\documentclass[aps,prx,reprint,amsmath,amssymb,superscriptaddress,longbibliography]{revtex4-2}
 
\usepackage[breaklinks=true]{hyperref}
\usepackage{fontenc}

\usepackage{graphicx}   
\usepackage{amsbsy} 
\usepackage{makecell}

\usepackage{natbib}
\setcitestyle{square, numbers}
\usepackage{verbatim}
\usepackage[OT1]{eulervm}
\usepackage{amsfonts}
\usepackage{amsmath}
\usepackage{amssymb}
\usepackage{amsthm}
\usepackage{mathtools}
\usepackage{braket}
\usepackage{xcolor}
\usepackage{bm}
\usepackage{multirow}
\usepackage{braket}
 
\usepackage[caption=false]{subfig}

\usepackage{bm}
\hypersetup{colorlinks=true, linkcolor=black}
\usepackage{bbm}

\begin{document}

\title{Encoding a bosonic mode into qubits for simulating boson-matter interactions}

\author{Maarten Stroeks}
\affiliation{Delft Institute of Applied Mathematics \& QuTech, TU Delft, Delft, The Netherlands}

\author{Jelena Mackeprang}
\affiliation{QuSoft and CWI, Science Park 123, 1098XG Amsterdam, The Netherlands}

\author{Yaroslav Herasymenko}
\affiliation{Perimeter Institute for Theoretical Physics, Waterloo, Ontario N2L 2Y5, Canada}

\author{Barbara Terhal}
\affiliation{Delft Institute of Applied Mathematics \& QuTech, TU Delft, Delft, The Netherlands}


\begin{abstract}
We explore the space-time cost of encoding of a bosonic mode into a set of qubits via various encodings, and their resilience to noise. In particular, we compare the performance of the unary Fock state encoding, an unexplored encoding of Fock states into Dicke states (``Dicke'' encoding), and the binary quadrature encoding. We do so by considering a displacement as well as a simulation of a Dicke model, coupling a bosonic mode to a set of atoms represented as qubits. Using classical simulations, we assess the accuracy and optimized space-time
costs in terms of number of qubits, circuit depth and CNOT count of a quantum simulation of the time dynamics using these encodings. Although our study is far from comprehensive, we find that the binary quadrature encoding is overall most favorable in terms of being very space-efficient and relatively time-efficient due to the use of the quantum Fourier transform. Our novel Dicke encoding leads to very shallow circuits, but pays in terms of using many more qubits. We probe the use of an H2-2 Quantinuum trapped-ion qubit device (56 qubits) for displacement and Dicke model simulation in the low boson-number regime, where we find that noise leads to visible deviations from ideal circuit dynamics.
\end{abstract}

\maketitle

\tableofcontents

\section{Introduction}

Quantum simulation of many-body fermionic and bosonic systems is a prime candidate for the use of a quantum computer~\cite{lloyd}. While fermions in second-quantized language fairly neatly map onto multi-qubit systems, bosonic degrees of freedom are mathematically modeled using an infinite-dimensional Hilbert space. As physical bosonic systems occupy only part of this phase-space and the resolution of tiny phase-space features is noise- and energy-limited, there is no telling whether `the infinite dimensionality' is just a convenient mathematical fiction or a physical truth. Even though states with a huge number of bosons are common place in classical physics (BEC, EM field), there are no realizations of genuinely quantum states like GKP grid or cat states~\cite{milul} with such huge boson numbers. For most practical purposes, one can model a bosonic mode as a qudit with a sufficiently large Hilbert space, or a bunch of qubits. 

However, this simple picture leaves questions. First, what is the {\em best} encoding of the finite-energy phase space with its {\em common} bosonic interactions, into qubits and their operations? Second, how does this translate to the logical level if we use quantum error correction? The latter topic is not the subject of this paper, but we muse about in Discussion Section~\ref{sec:discuss-logical}. 

What are common bosonic interactions? All of physics (field theories etc.) starts with fermion-boson interactions which are linear or quadratic in bosonic and fermionic creation and annihilation operators. Higher-order terms only come up effectively through these basic interactions.
Such linear and quadratic interactions for bosons and fermions {\em separately (and thus uncoupled)} are Gaussian and entirely fall within the domain of classical simulatibility. 

While it may be possible (but not clear) that one can leverage these facts to construct very efficient bosonic+fermionic encodings, in this paper we instead investigate and compare properties of known boson-into-qubits encodings and introduce one new one. Our comparison focuses on number of qubits needed, depth and CNOT count for some standard benchmarks, namely a displacement and the simulation of a Dicke model. 

Encoding a boson using a discretized x-quadrature of a bosonic mode (binary quadrature encoding) was considered in e.g. Refs.~\cite{JLP,dig:boson,Somma1Dsystems,QuantumHermiteTransform, Motlagh_2025,lu2026}. Encoding a boson using Fock states represented in unary or binary or some variants thereof was considered in e.g.~\cite{Sawaya, dig:boson, crane2024+, peng2025, somma+:qsimul, majland2025ft}. Effects of truncating an infinite-dimensional Hilbert space for quantum computer simulations were analyzed in~\cite{Tong2022provablyaccurate}.

In the next Section~\ref{sec:encodingabosonicmode} we review these known encodings and introduce the Dicke encoding. We then discuss the space-time cost of realizing a displacement using these encodings. In Section~\ref{sec:Dicketimeevo} we extend our study to benchmark the space-time cost for simulating the time-evolution of the Dicke model, describing some choices for circuit optimization. To ground our expectations, in Section~\ref{sec:h22} we run the simulation of a displacement and time evolution of the Dicke model on a 56-qubit trapped-ion qubit device H2-2, comparing unary or Dicke encoding. We finish the paper with an extensive Discussion Section~\ref{sec:discuss}, and stuff some extra details in a few Appendices.

\section{Encoding a bosonic mode into a set of qubits}
\label{sec:encodingabosonicmode}

This section surveys the existing encodings and our novel `super-inefficient' Dicke encoding to encode a bosonic mode with creation operator $a^{\dagger}$, $[a,a^{\dagger}]=I$, into a set of qubits. Fock states of the bosonic mode are denoted as $\ket{m}$, $a^{\dagger}\ket{m}=\sqrt{m+1}\ket{m+1}$ with $m=0,1,\ldots, \infty$. Throughout, we use $N_b$ qubits to represent the bosonic mode and subscript $_E$ to denote encoded operators. We will use a displacement
\begin{equation} \label{eq:displacement}
D(\alpha) = \exp(\alpha a^{\dagger} - \alpha^{*}a),
\end{equation}
for $\alpha \in \mathbb{C}$, to benchmark some of the features of these encodings in Section \ref{sec:displacement_encoding}.

In encodings in which only a subspace of the total qubit space is used, such as the unary Fock and Dicke encoding, it is clear that there are multiple ways of representing an operator $f(a,a^{\dagger})$ since it only has to be the correct operator on the subspace. One then wishes to choose the simplest representation of such operator. In all encodings, we wish to encode a small part of phase-space centered around the vacuum with $\langle a^{\dagger} a\rangle < N_{\rm cutoff}$.

\begin{figure*}[htb]
    \centering
    \subfloat[Unary encoding.\label{Unary encoding.}]{
        \includegraphics[width=0.325\textwidth,height=0.25\textheight,keepaspectratio]{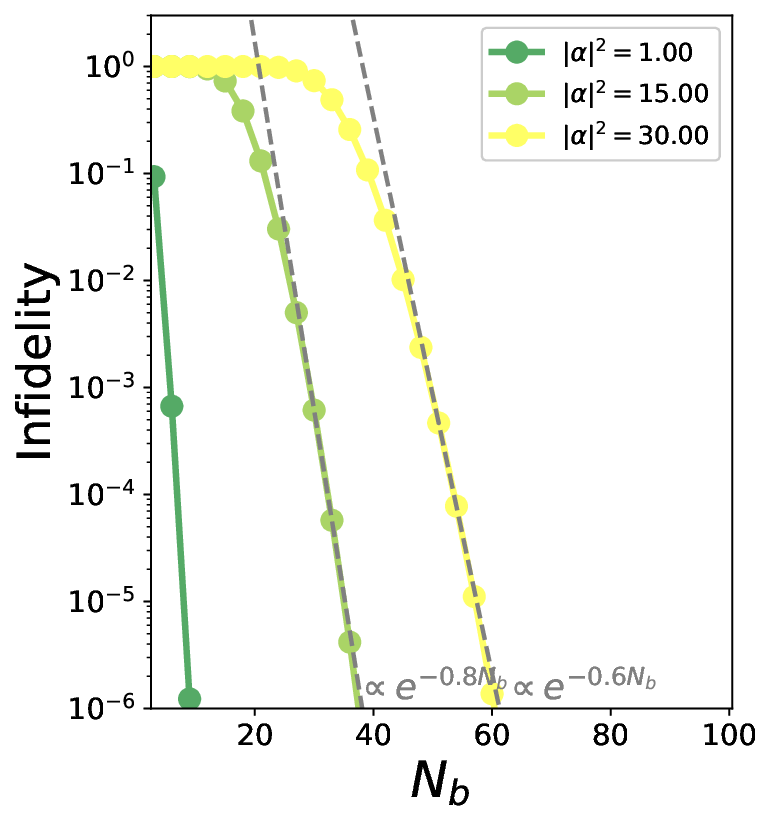}
    }
    \hfill
    \subfloat[Dicke encoding.\label{Dicke encoding.}]{
        \includegraphics[width=0.325\textwidth,height=0.25\textheight,keepaspectratio]{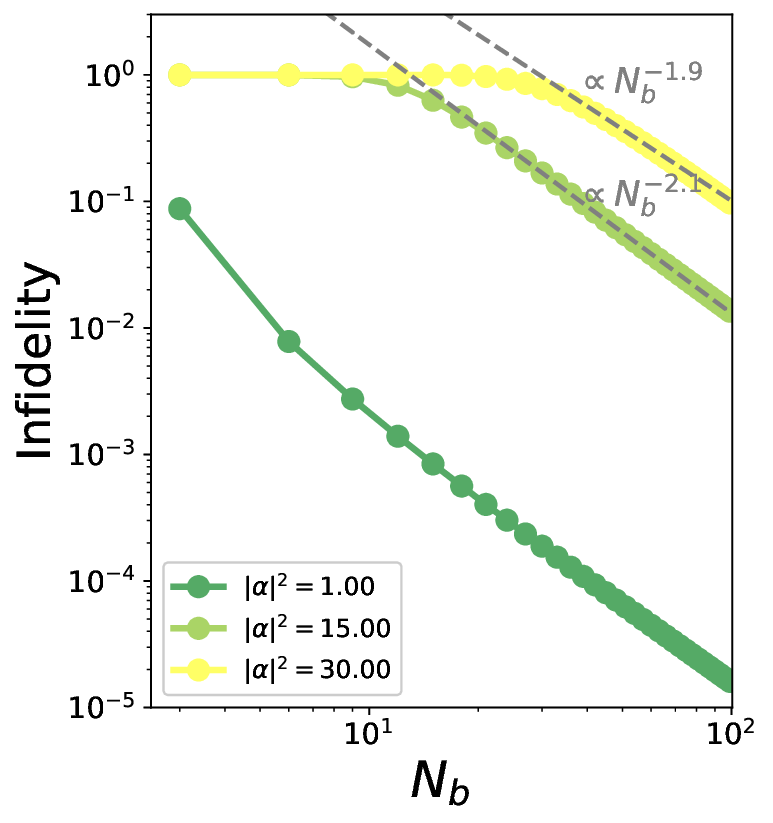}
    }
    \hfill
    \subfloat[Binary-x encoding.\label{Binary-x encoding.}]{
        \includegraphics[width=0.325\textwidth,height=0.25\textheight,keepaspectratio]{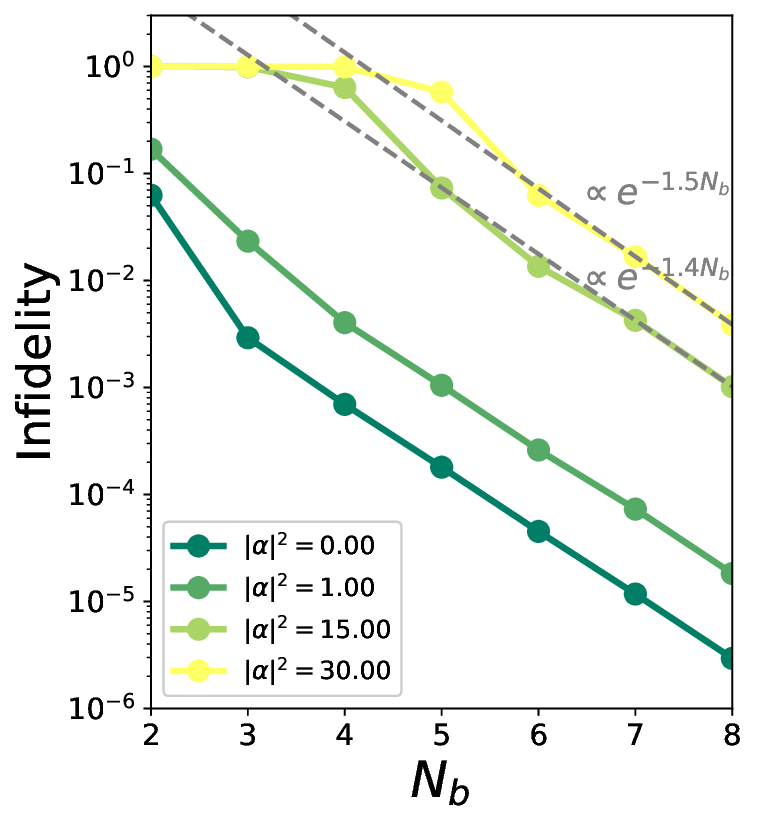}
    }
    \caption{Encoding error expressed as the infidelity of encoded displaced vacuum states w.r.t. the exact displaced vacuum state $\ket{\alpha}$. The infidelity w.r.t. the exact displaced vacuum state is approximated by $1-\bigl\rvert\langle\psi_{E}|\:\Psi_{N_b'}\rangle\bigr\rvert^2$, where $\ket{\psi_{E}}$ is the encoded $N_b$-qubit displaced vacuum \textit{extended} to $N_b'$ ($\:>N_b$) qubits and $\ket{\:\Psi_{N_b'}}$ is the $N_b'$-qubit displaced vacuum state, where $N_b'$ is taken large enough for $1-\bigl\rvert\langle\psi_{E}|\:\Psi_{N_b'}\rangle\bigr\rvert^2$ to be converged for all practical purposes. The \textit{extension} procedure corresponds to appending $\ket{\psi_{E}}$ with zeros in the unary and Dicke encoding, and to a linearly interpolating $\ket{\psi_{E}}$ in the binary-x encoding. For each $|\alpha|^{2}$ curve, we have calculated the infidelity by averaging over the phase of $\alpha$.}
    \label{fig:encoding_errors}
\end{figure*}

\subsection{Unary Fock encoding}
\label{sec:unifock}
The unary encoding maps a Fock state $\ket{m}$ to $\ket{m_E}=\ket{0\ldots0_{m-1}\:1_m\: 0_{m+1}\ldots 0}$ with $m\in\{0,1,\ldots,N_b-1\}$. To encode all Fock states less than $N_{\rm cutoff}$, one thus uses $N_{\rm cutoff}$ qubits. The encoded annihilation operator equals
 \begin{equation}
    a \: \to \: a_E := \sum_{j=0}^{N_b-2}\sqrt{j+1}\:\sigma_{j+1}^{-}\sigma^{+}_{j}.
\end{equation}
Thus $a_E^{\dagger}$ exactly represents the action of $a^{\dagger}$ on any Fock state $\ket{m < N_b-1}$, while it fails on $\ket{m=N_b-1}$, i.e.  $a_E^{\dagger}\ket{m_E=N_b-1}=0$. When restricting to the unary encoding subspace, we have
\begin{equation}
 a_E^{\dagger}a_E =\sum_{j=0}^{N_{b}-1}\frac{j}{2}(\mathbbm{1}-Z_{j}), \label{eq:num_operator_unary}   
\end{equation} representing $a^{\dagger}a$ exactly on this subspace. Note that the preparation of the Fock state $\ket{m_E}$ is trivial using a single-qubit $X$ gate on qubit $m$, starting from $\ket{00\ldots 0}$.



\subsection{Binary Fock encoding}
\label{sec:binfock}
In the binary Fock encoding one simply encodes Fock states in binary with $N_b=\lceil \log_2 N_{\rm cutoff}\rceil$, leading to a qubit overhead savings as compared to unary encoding. Variants of binary encoding, such as Gray encoding or hybrid binary-unary also exist~\cite{Sawaya}. We have refrained from including the binary Fock encoding in our benchmarks due to the circuit size and depth overhead of implementing displacements $D(\alpha)$, see the discussion in Section~\ref{sec:binaryF}.

\subsection{Dicke encoding}
\label{sec:bindicke}
In the Dicke encoding we take the permutation-symmetric Dicke states to represent Fock states, that is,
\begin{equation}
    \ket{m} \rightarrow \ket{m_E}= \frac{1}{\sqrt{\binom{N_b}{m}}}\sum_{x \in \{0,1\}^{N_{b}}\:\text{s.t.}\:|x|=m}\ket{x}.
\end{equation}
The annihilation operator is represented as
\begin{equation}
    a \: \to \: a_E := \frac{1}{\sqrt{N_b}}\sum_{j=0}^{N_b-1}\sigma^{-}_{j},
    \label{eq:anni-Dicke} 
\end{equation}
so that 
\begin{align}
    a_E \ket{m_E} =&\: f(m-1)\sqrt{m}\ket{m-1_E}, \label{eq:def_a_dicke}\\ \nonumber  
    a_E^{\dagger}a_E\ket{m_E} =&\: f^2(m-1)\:m\ket{m_E},
\end{align}
where $f(m) := \sqrt{\frac{N_b - m}{N_b}}\approx 1$ when $m \ll N_b$. One can view the Dicke encoding as a Holstein-Primakov conversion from a bosonic mode to a large spin represented by qubits, taken at the leading order in inverse system size. Hence, $a_E^{\dagger}$ represents the bosonic creation operator in a deformed way, not only affecting the high-excitation Fock states as in the unary encoding. One thus requires $N_b\gg N_{\rm cutoff}$ to have a small encoding error, making the encoding `super-inefficient'. This space inefficiency comes with a time efficiency advantage, i.e. displacements require a circuit of single-qubit gates of depth-1 as will be discussed in Section \ref{sec:displace_circuit}. On the Dicke encoding subspace, we can represent $a^{\dagger} a$ by using the simple expression
\begin{align}
    a^{\dagger}a \rightarrow  \frac{1}{2}\sum_{j=0}^{N_b-1}(\mathbbm{1}-Z_{j}).
    \label{eq:reps-n}
\end{align}
Note that this is only approximately equal to $a_E^{\dagger}a_E$ for finite $N_b$. 

For this encoding, preparing a Fock state amounts to preparing a Dicke state, for which high-probability-output circuits have been developed with depth $\text{polylog}(N_b)$, using $\text{polylog}(N_b)$ ancillary qubits~\cite{ChildsDicke}. Note that the complexity of preparing a Dicke state is lower for low excitation number $m$ (the same for low $N_b-m$). The vacuum state is simply $\ket{00\ldots 0}$. For example, one can take a single ancilla qubit in $\ket{1}$ and symmetrically couple it, via a flip-flop interaction ($\sigma_i^+\sigma_j^-+h.c.$), to all $N_b$ qubits, in order to add one more excitation to the bosonic qubit register $m\rightarrow m+1$ and leave the ancilla qubit in $\ket{0}$~\cite{Hakoshima_2020}. The interaction with this ancilla qubit would then need to be further Trotterized to be decomposed into two-qubit gates. 

As an aside, note that one can also encode a bosonic mode into a set of $2N_b$ fermionic modes (with annihilation operators $c_i$) using this encoding by choosing 
\begin{align}
    a \rightarrow \frac{1}{\sqrt{N_b}}\sum_{i=0}^{N_b-1}c_i c_{i+N_b},
\end{align}
so that the bosonic algebra is approximately reproduced when $N_b\gg N_{\rm cutoff}$.


\subsection{Binary quadrature encoding}
\label{sec:binquad}

Here we encode a bosonic mode into a discretized quadrature, for example the $x$-quadrature, using $N_b = \lceil \log_{2}(L) \rceil$ qubits, where $L$ is a discretization parameter. In this paper we will consider binary-x quadrature encoding, but for some Hamiltonian simulations the choice of quadrature could be optimized differently of course.
One represents the value of the position quadrature $x\in \mathbb{R}$ with $\hat{x} = \frac{1}{\sqrt{2}}(a^{\dagger} + a)$ by $x_E=\sqrt{\frac{2\pi}{L}}(-L/2+j)$ with $j=0,\ldots,L-1$, so the encoding is the bitstring $\ket{j}$ with $\hat{j}=\sum_{i=0}^{N_b-1} 2^{i} \frac{1}{2}(I-Z_i)$. For simplicity, we will choose $L=2^{N_b}$, and this gives
\begin{align}
  x \rightarrow x_E :=&\: \sqrt{\frac{2\pi}{2^{N_b}}}\left(-\frac{1}{2}-\sum_{i=0}^{N_b-1} 2^{i-1}Z_i\right).
  \label{eq:xbin}
  \end{align}
It is straightforward to show that (for $N_b\geq 2$) one can map $\hat{p}= \frac{i}{\sqrt{2}}(a^{\dagger}-a) \rightarrow p_E$ defined as
\begin{equation}
    p_E = (Z_0\:QFT^{\dagger}\:Z_0) \:x_E\: (Z_0\:QFT\:Z_0).
    \label{eq:pbin}
\end{equation}
Here, the $QFT$ denotes the $N_b$-qubit quantum Fourier transform s.t. $\bra{j} QFT \ket{k} = \frac{1}{\sqrt{L}}\exp(i2\pi jk/L)$ with $j,k\in 0,1,\ldots, 2^{N_b}-1$.
The encoded annihilation operator follows as
\begin{equation}
    a \: \to \: a_E := \frac{1}{\sqrt{2}}(x_E+ip_E).
\end{equation}
In principle, the encoded boson number is given by $a_E^{\dagger}a_E = \frac{1}{2}(x_E^{2} + p_E^{2} + i[x_E,p_E])$, where $[x_E,p_E]\neq iI$ due to the encoding error. In practice, the encoding error is incurred by decomposing the time-dynamics of $a_E^{\dagger}a_E$ using fast-forwarding or Trotterization into dynamics with $\sim x_E^2$ and $\sim p_E^2$, see Section \ref{sec:freeH}. 

In the binary-x encoding, Fock states $\ket{m}$ are the Hermite functions $\psi_m(x)$. The vacuum state is the Gaussian function $\psi_0(x)=\pi^{-1/4}e^{-x^2/2}$ whose preparation we discuss in Appendix \ref{sec:gauss}.
The preparation of higher discretized Hermite functions has been studied in Ref.~\cite{QuantumHermiteTransform}, where it is shown that their preparation with constant error (in $l_2$-norm) can be done with circuits with $\text{poly}(N_b)$ CNOT count.

\subsection{Displacement: encoding error and cost}\label{sec:displacement_encoding}

\subsubsection{Displacement: encoding error}

 All three encodings make an encoding error when representing a displacement $D(\alpha)\ket{m=0}=\ket{\alpha}$ due to representing only part of phase space. A first question is how this encoding error decreases as a function of $N_b$. Naturally, the encoding error increases with $|\alpha|^2$ as a coherent state $\ket{\alpha}$ has $\langle a^{\dagger}a\rangle=|\alpha|^2$ which should be below the number of Fock states which are represented. Fig.~\ref{fig:encoding_errors} displays some numerical results; note that this encoding error does not include circuit realization errors, e.g. Trotter errors. For all encodings, $N_b$ should be chosen sufficiently large as compared to $|\alpha|^2$ for the infidelity to be on the large $N_b$ scaling curve. This curve decays exponentially in $N_b$ for the unary and binary-x encoding, while in the Dicke encoding we observe a polynomial decay. This could be explained by the fact that for unary Fock encoding the bosonic algebra is only misrepresented at high excitation number $m$. For binary-x encoding the error can be expected to scale as $1/{\rm poly}(L)$ which is exponentially small in $N_b$.
 For Dicke encoding the bosonic algebra is misrepresented throughout the whole Hilbert space due to the pre-factor $f(m-1)$ in Eq.~(\ref{eq:def_a_dicke}). Note that for the Dicke encoding, the value of $N_b$ needed to suppress the encoding error is also considerably larger than for either unary Fock, or compare the very low $N_b$ numbers for the binary-x encoding. Lastly, we note that the vacuum state does not suffer from an encoding error in the unary and Dicke encodings, while in the binary-x encoding it \textit{is} an approximate (Gaussian) state represented by the dark green curve in Fig.~\ref{fig:encoding_errors}(c).

One expects that binary Fock encoding would perform similarly as binary $x$-encoding in having an exponentially-decreasing infidelity, but see Section~\ref{sec:binaryF}. 
 
Not surprisingly, we observe that the phase of $\alpha$ affects the encoding error in the binary-x encoding, hence we take an average (obviously, it has no relevance in the unary and Dicke encoding).

While for unary and Dicke encoding, the relation between $N_b$ and $N_{\rm cutoff}$ for a given infidelity is quite direct and approximately linear, arguments in Ref.~\cite{dig:boson} show that for the binary quadrature encoding one has $2^{N_b}\sim N_{\rm cutoff}$ for a targeted fixed infidelity. Note that if want the encoded circuits for the binary quadrature encoding to be efficient in $N_{\rm cutoff}$, we are then allowed circuits with exponential complexity in $N_b$.

\subsubsection{Displacement: circuit realization}
\label{sec:displace_circuit}
\begin{table*}[htb]
\centering
\renewcommand{\arraystretch}{1.3}
\resizebox{\textwidth}{!}{%
\begin{tabular}{||l|c|c|c|c||}
\hline 
Encoding & Circuit type & 
\makecell{Circuit depth} & 
\hspace{0.2cm}\makecell{CNOT count}\hspace{0.2cm} &
\hspace{0.2cm}\makecell{Trotter error $\epsilon_T$}\hspace{0.2cm} \\
\hline\hline
Dicke      & 1-qubit gates & 1 & 0 & 0  \\
Unary Fock  & 2-qubit gates & $18\: n_T$ & $4 (N_b-1) n_T$ & $O(|\alpha|^2 N_b/n_{T})$  \\
Binary quadrature & \hspace{0.2cm}QFT + 1-qubit gates\hspace{0.2cm} & \hspace{0.2cm}$12N_b-11 \:^*$\hspace{0.2cm} & \hspace{0.2cm}$2N_b(N_b-1) \:^*$\hspace{0.2cm} & 0 \\
\hline
\end{tabular}%
}
\caption{Comparison between encodings in terms of cost (circuit depth -- in terms of CNOTs and single-qubit gates -- and CNOT count) for implementing a displacement $D(\alpha)$ using $N_b$ qubits to represent the bosonic mode. The Trotter error is defined as $||D_E(\alpha)-\tilde{D}_E(\alpha)||_E\leq \epsilon_T$, where $D_E(\alpha)$ is the \textit{encoded} displacement operation, $\tilde{D}_E(\alpha)$ is the Trotterized, encoded, circuit using $n_T$ Trotter layers and $\lvert\lvert \cdot \rvert\rvert_{E}$ denotes the spectral norm on the encoded subspace. $^*$This CNOT count and depth could be improved by considering QFT implementations of reduced depth/size~\cite{CW:QFT}, but such optimization is not relevant for the small $N_b$ studied here.} 
\label{tab:displacement-encoding-comparison}
\end{table*}

Given the encoding of the operator $a\rightarrow a_E$ as described in the previous sections, let us discuss the space-time cost of the circuit realization of a displacement $D(\alpha)$, see the summary in Table~\ref{tab:displacement-encoding-comparison}. 
Here, a clear advantage of the Dicke encoding is observed: since $a_{E}$ is a sum of single-qubit gates on the bosonic qubits, a displacement can be executed using only single-qubit gates, i.e.
\begin{multline}
    D_E(\alpha) = \bigotimes_{j=0}^{N_b-1} \exp\Big[\frac{1}{\sqrt{N_b}} i\big( -Re(\alpha)Y_j + Im(\alpha)X_j \big) \Big]. \label{eq:displacement_dicke}
\end{multline}

A displacement in the unary encoding is given as
\begin{multline}\label{eq:displacement_unary}
    D_E(\alpha) = \exp\big[ \sum_{j=0}^{N_b-2}\sqrt{j+1}\big(\alpha\sigma_{j+1}^{+}\sigma_{j}^{-} - \alpha^{*}\sigma_{j+1}^{-}\sigma_{j}^{+}\big) \big],
\end{multline} 
requiring Trotterization to be broken up into two-qubit gates, and hence incurring a Trotter error. Indeed, we can Trotterize to a brick-work circuit
\begin{equation}\label{eq:displacement_unary_trotter}
D_E(\alpha)\approx \left(e^{-iH^e(\alpha/n_T)}e^{-iH^o(\alpha/n_T)}\right)^{n_T}.
\end{equation}
with Trotter parameter $n_T$ and 
\begin{align}
 H^e(\alpha)&=i\sum_{j=0,\;j\;{\rm even}}^{N_b-2} \sqrt{j+1} \big(\alpha\sigma_{j+1}^{+}\sigma_{j}^{-} - \alpha^{*}\sigma_{j+1}^{-}\sigma_{j}^{+}\big), \label{eq:displacement_unary_He} \\
H^o(\alpha)&=i\sum_{j=1, \;j\;{\rm odd}}^{N_b-2} \sqrt{j+1} \big(\alpha\sigma_{j+1}^{+}\sigma_{j}^{-} - \alpha^{*}\sigma_{j+1}^{-}\sigma_{j}^{+}\big). \label{eq:displacement_unary_Ho}
\end{align}
The induced Trotter error of this decomposition, given in the last column of Table \ref{tab:displacement-encoding-comparison}, is explicitly bounded in Appendix~\ref{app:Trotter_derivation}. 

As for a displacement in the binary-$x$ encoding, one can first write
\begin{multline}
    D(\alpha) = e^{-i Re(\alpha)Im(\alpha)}e^{i\sqrt{2}{\rm Im}(\alpha)x}e^{-i\sqrt{2}Re(\alpha)p},
    \label{eq:dis_x}
\end{multline}
and then replace $x\rightarrow x_E$ and $p\rightarrow p_E$ and using the QFT, one gets
\begin{multline}
    D_E(\alpha) = e^{-i Re(\alpha)Im(\alpha)}e^{i\sqrt{2}Im(\alpha)\:x_E} \times \\  (Z_0\:QFT^{\dagger}\:Z_0) e^{-i\sqrt{2}Re(\alpha)\:x_E} (Z_0\:QFT\:Z_0).
\label{eq:binaryxdisplacement}
\end{multline}
The operation $\exp(i\theta x_E)$ for any $\theta$ can be implemented as a product of single-qubit gates using the representation of $x_E$ in Eq.~\eqref{eq:xbin}. Thus the cost of a displacement in the binary-x encoding is that of single-qubit gates plus two executions of the QFT for which there are standard circuits~\cite{book:NC, CW:QFT}, and no further Trotter error. A similar binary-x displacement construction is given in~\cite{lu2026}. Note that a displacement for purely imaginary $\alpha$ does not require the QFT and hence only uses a layer of single-qubit gates. 

\subsubsection{Comparison with binary Fock encoding}
\label{sec:binaryF}

\begin{figure}[htb]
\centering
\includegraphics[width=0.48\textwidth,height=0.32\textheight,keepaspectratio]{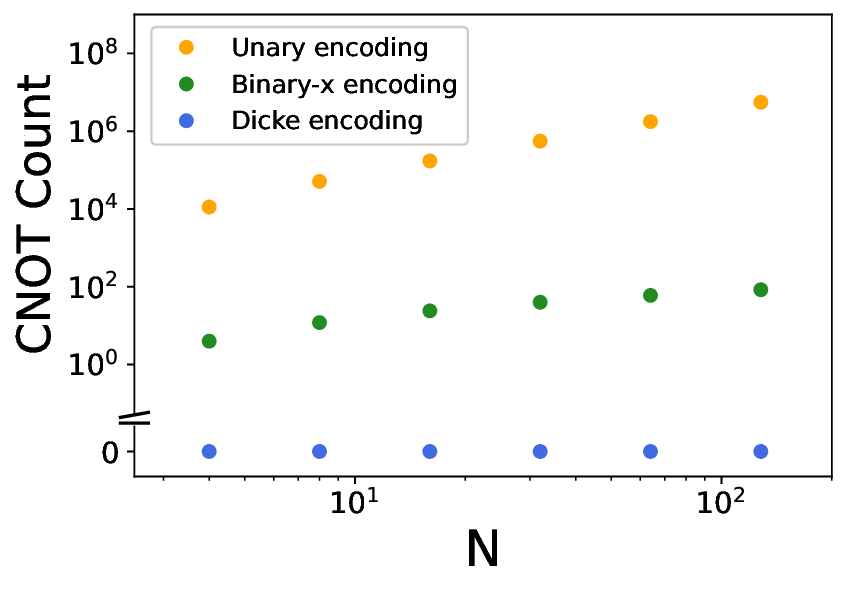}
\caption{CNOT count required to achieve an error $\bigl\lvert\bigl\lvert D_{E}(\alpha) - \tilde{D}_{E}(\alpha) \bigr\rvert\bigr\rvert_{E} \leq 10^{-3}$, with $\tilde{D}_{E}(\alpha)$ the circuit implementation of the encoded displacement $D_{E}(\alpha)$ using a certain number CNOT gates, and $\lvert\lvert \cdot \rvert\rvert_{E}$ denoting the spectral norm on the encoded subspace. We take $|\alpha| = 1$ and the plotted CNOT counts are the worst-case CNOT counts over the phase of $\alpha$. Here $N = N_b$ for the unary Fock encoding, $N = N_b+1$ for the Dicke encoding and $N = 2^{N_b}$ for the binary-x encoding, hence $N$ can be understood as a cutoff. Note that we do not consider the error imposed by the encoding itself here. }
\label{fig:CNOTcount_displacements}
\end{figure}

The binary Fock encoding may be the most qubit-efficient encoding. However, as observed in Section VIII in~\cite{liuhybrid2026}, it has some major drawbacks in terms of cost to implement a displacement, perhaps precisely due to such compressed representation. 
In this encoding, the operator $i(\alpha a^{\dagger}_E-\alpha^* a^{\dagger}_E)$ is a sparse off-diagonal matrix containing $\sqrt{.}$ factors. According to~\cite{liuhybrid2026}, the computation of the $\sqrt{.}$ function of the integer $m$ represented in binary is cumbersome, requiring a deep and CNOT-intensive quantum circuit.  Concretely, to see the difference with the other three encodings, we consider the following. Fig.~41 in Ref.~\cite{liuhybrid2026} estimates a CNOT count for a displacement $D(\alpha)$ with $|\alpha|=1$ with targeted error $\delta=10^{-3}$ as compared to the exact $D_E(\alpha)$ \footnote{Hence, this $\delta$ error does not include the encoding error itself.} given a cutoff $N=2^{N_b}$. This plot can be compared to our Fig.~\ref{fig:CNOTcount_displacements} which plots the CNOT count for the encodings studied in this paper in an identical manner as a function of $N$. Note that Fig.~41 is an upper bound on the CNOT count that is obtained if we set the error and $N$ based on semi-analytical estimates in the analysis of~\cite{liuhybrid2026}, hence possibly overestimating the CNOT count. We observe that the CNOT counts for binary Fock encoding, even for the brute force method in Fig. 41, lie orders of magnitude above what binary-x encoding achieves in our corresponding Fig.~\ref{fig:CNOTcount_displacements}. Of course, for the Dicke encoding the CNOT count in Fig.~\ref{fig:CNOTcount_displacements} is zero at all $N$. Further comparisons between the binary Fock and binary-x encodings for implementing displacements and other operations are given in~\cite{lu2026}. 

One can change from the binary Fock to the binary-x encoding via the Hermite transform in Ref.~\cite{QuantumHermiteTransform} (Theorem 19): one maps the Fock state $\ket{m}$ onto its representation in $x$-space, the Hermite function $\psi_m(x)$. According to the Theorem this does incur some poly overhead in the number of qubits, showing that these encodings are not `cost-equivalent' if one focuses on small $N_b$ optimizations (so including polynomial and constant factors).

\subsubsection{Comparison between encodings, including circuit depth and CNOT count}
Fig.~\ref{fig:CNOTcount_displacements} shows that the Dicke and binary-x encoding do well in terms of CNOT count when comparing the encoded circuit with the Trotterized encoded circuit, while Fig.~\ref{fig:encoding_errors} shows that the binary-x encoding does particularly well in terms of space-overhead if one only considers the encoding error. Thus to make a comprehensive comparison, we can combine the data from these Figures as follows. One can bound 
\begin{align}
    &||\ket{\alpha}-\tilde{D}_E(\alpha) \ket{\tilde{0}_E}|| \notag \\ &\leq||\ket{\alpha}-\ket{\alpha_E}||+||\ket{\alpha_E}-\tilde{D}_E(\alpha)\ket{0_E}||\notag \\ &\hspace{4.0cm}+||\tilde{D}_E(\alpha)\ket{0_E} - \tilde{D}_E(\alpha)\ket{\tilde{0}_E}|| \notag \\ &\lessapprox\sqrt{2(1-|\braket{\alpha|\alpha_{E}}|^2)} +||D_E(\alpha)-\tilde{D}_E(\alpha)||\notag \\ &\hspace{4.0cm}+\sqrt{2(1-|\braket{0_{E}|\tilde{0}_E}|^2)}, 
    \label{eq:upperb}
\end{align}
where $1-|\braket{\alpha|\alpha_E}|^2$ is the infidelity in Fig.~\ref{fig:encoding_errors} \footnote{$\ket{\psi_E}$ has to be understood as the encoded version of a state $\ket{\psi}$ represented in the full bosonic Hilbert space without any cutoff}. Note that $|\braket{0_{E}|\tilde{0}_E}|^2 = 1$ for the unary and Dicke encoding, and for the binary-x encoding we use the preparation method from Appendix~\ref{sec:gauss} -- where we take the minimum-depth preparation circuit achieving the desired accuracy $\epsilon_*$, for each $N_b$. 

\begin{figure}[htb]
    \centering
    \subfloat[Circuit depth.\label{}]{
        \includegraphics[width=0.42\textwidth,height=0.45\textheight,keepaspectratio]{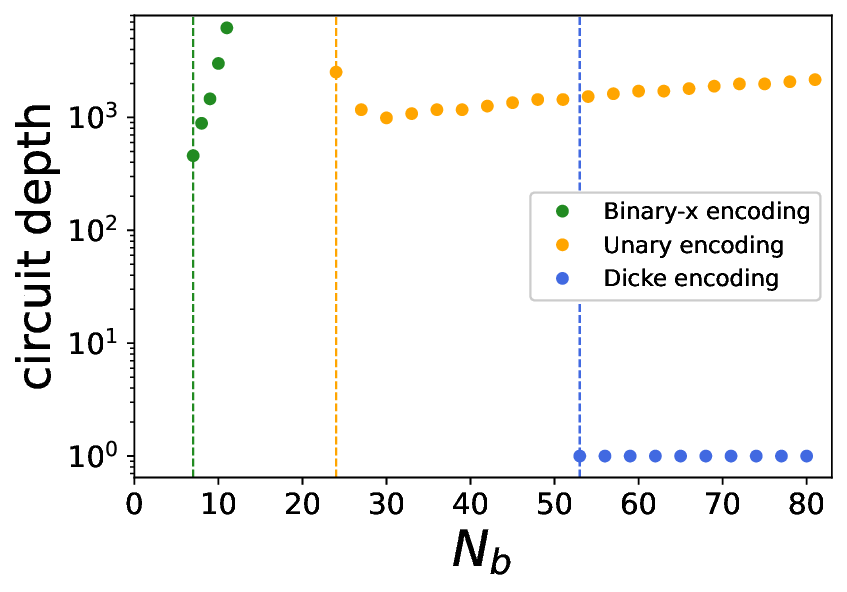}
    }
    \vfill
    \subfloat[CNOT count.\label{}]{
        \includegraphics[width=0.428\textwidth,height=0.45\textheight,keepaspectratio]{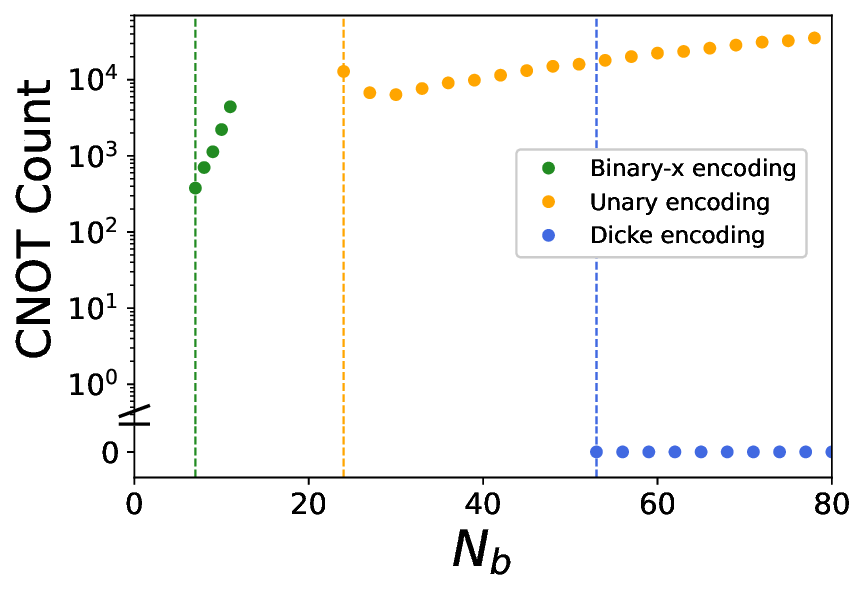}
    }
    \caption{Circuit depth and CNOT count required to achieve a given error $\epsilon_*$ on the displacement operator $D(\alpha)$ -- for \textit{any} $\alpha$ s.t. $|\alpha|^2 = 15$ -- as a function of $N_b$. To simplify our analysis, we say that error $\epsilon_*$ is attained if the upper bound in Eq.~\eqref{eq:upperb} is smaller than $\epsilon_*$. We take $\epsilon_* = 0.3$. The vertical lines indicate the lowest $N_b$ values with such small error $\epsilon_{*}$. We simulate up to $N_b = 10$ for the binary quadrature encoding due to memory constraints. }
    \label{fig:displacement_depths}
\end{figure}

In Fig.~\ref{fig:displacement_depths} we compare the encodings from the point of view of total circuit depth and CNOT count for implementing a displacement ---taking into account the full pipeline of encoding error, approximate preparation of the encoded vacuum state, and circuit implementation of the displacement to achieve an error $\epsilon_*$ on the bound in Eq.~\eqref{eq:upperb}. Clearly, which encoding is optimal depends on which resource is more abundant -- circuit depth/CNOT count or qubit count. Although the binary-x encoding is very space-efficient, the Dicke encoding only requires depth-one circuits and no CNOTs. 


\subsection{Free bosonic dynamics in the three encodings}
\label{sec:freeH}
Free evolution of the bosonic mode seems an important primitive in any bosonic dynamics (see however also Discussion Section \ref{sec:rot_frame}). A phase-space rotation $R(\theta)=\exp(i\theta a^{\dagger}a)$ has the simplest expression in the unary Fock and Dicke encodings where it is a product of single-qubit gates due to Eq.~(\ref{eq:num_operator_unary}) and Eq.~(\ref{eq:reps-n}). For the binary Fock encoding the dynamics is generated by multi-qubit $Z$ gates instead of single-qubit $Z$ rotations, leading to a depth and CNOT count overhead. For the binary-x encoding we can use fast-forwarding of the harmonic oscillator Hamiltonian $H=\omega (a^{\dagger} a+\frac{1}{2})= \frac{\omega}{2}(x^2+p^2)$, i.e. one can write $\exp(-i H t)$, see Theorem 3 in~\cite{QuantumHermiteTransform}, as 
\begin{align}
    U=e^{-i H t}= e^{-i a(t) x^2} e^{-i b(t) p^2} e^{-i a(t) x^2},
    \label{eq:ff}
\end{align}
$a(t)=\tan(t/2)/2$ and $b(t)=\sin(t)/2$. As $a(t)$ diverges at $t=(2k+1)\pi$, we use this formula only when $t\mod 2\pi \in [-\pi/2,\pi/2]$, otherwise one splits the time-evolution in half and uses the formula twice for $t/2$.

In the binary-x encoding, using Eq.~(\ref{eq:pbin}), Eq.~\eqref{eq:ff} can be realized as
\begin{align*}
U_E= e^{-i a(t) x_E^2}  (Z_0\:QFT^{\dagger}\:Z_0) e^{-i b(t) x_E^2} (Z_0\:QFT\:Z_0) e^{-i a(t) x_E^2}.
\end{align*}
Using Eq.~\eqref{eq:xbin}, 
the transformation $e^{-i a x_E^2}$, for any $a$, can be implemented by a $(1 + N_b)$-depth circuit using single-qubit $Z$ rotations and two-qubit $Z$-like (diagonal) gates. This holds since the all-to-all $\binom{N_b}{2}$ two-qubit diagonal gates can be implemented with depth $N_b$, as one can edge color an $N_b$-vertex complete graph using $N_b$ colors (Vizing's theorem). Note that the implementation of free bosonic evolution then requires at most 4 QFTs and a depth $5(1+N_b)$ circuit of two-qubit gates (in the worst case when we use the formula in Eq.~\eqref{eq:ff} twice).
In the cost estimates for the QFT in Table~\ref{tab:displacement-encoding-comparison}, and Table~\ref{tab:encoding-comparison} in Appendix~\ref{app:circuitdecompDicke}, we use the implementation of the $N_b$-qubit QFT of depth $6N_b-7$ (in terms of CNOTs and single-qubit gates) and CNOT count $N_b(N_b-1)$, see Ref.~\cite{parallelizedQFT}. Note that the CNOT count and depth could be improved by considering QFT implementations of reduced depth/size which make an additional approximation~\cite{CW:QFT}, but such optimization is not relevant for the small $N_b$ studied here.

The circuit realizations of the free bosonic evolutions will come into play when considering the simulation of a Dicke Hamiltonian.

\subsection{Other bosonic interactions}

Naturally, the cost of implementing a higher-order nonlinearity such as $(a^{\dagger}a)^2$ is higher than that of $a^{\dagger}a$ itself. For $Z$-like $a^{\dagger}a$ (unary and Dicke), one gets higher-weight $Z$-like (diagonal) gates. For the binary encoding, the nonlinear Hamiltonian $\sim (x^2+p^2)^2$ which, due to being non-diagonal, will incur a large space-time cost.

More generally, for parts of an evolution where the dynamics is generated by elements of a small Lie algebra (such as free fermions or free bosons) one can use fast-forwarding techniques~\cite{Gu_2021}.


\section{Time evolution with the Dicke Hamiltonian}\label{sec:Dicketimeevo}

\subsection{Dicke Hamiltonian}

Beyond the displacement operator considered in Sec.~\ref{sec:displacement_encoding}, we would like to benchmark the encodings for simulating the time-dynamics of a Hamiltonian of some physical interest. A natural candidate is the Dicke Hamiltonian~\cite{DickeModel}, which couples $N_a$ (``atom'') qubit degrees of freedom to a single bosonic mode (a ``photon''),
\begin{equation}\label{eq:H_dicke}
    H = \omega\, (a^{\dagger}a+\frac{1}{2}) - \frac{\omega_0}{2}\sum_{j=1}^{N_a} Z_{j} + \frac{g}{\sqrt{N_a}}\sum_{j=1}^{N_a} X_j \otimes (a+a^{\dagger}),
\end{equation}
with $\omega,\omega_0,g\geq 0$. We define, for later reference, 
\begin{align}
    H_b=\omega\, (a^{\dagger}a+\frac{1}{2}), \\
    H_a= - \frac{\omega_0}{2}\sum_{j=1}^{N_a} Z_{j}, \\  H_{ab}=\frac{g}{\sqrt{N_a}}\sum_{j=1}^{N_a} X_j \otimes (a+a^{\dagger}),
    \label{defs:Hs}
\end{align}
with corresponding encoded versions $H_{b,E},H_{a,E},H_{ab,E}$.

One can generalize this model to the disordered `inhomogeneous' Dicke model where each atom has a random frequency $\omega_j$ and coupling parameter $g_j$. For such disordered model, one expects classical simulations to scale superpolynomially in $N_a$. So in this case a quantum simulation may provide a quantum advantage. Trotter-based quantum simulation of the disordered model is clearly no more costly than that of the `clean' Dicke model in Eq.~\eqref{eq:H_dicke}. Due to the permutation symmetry between the atom-qubits in this clean model, the dynamics of $H$ can be classically simulated in the $N_a+1$-dimensional subspace (often modeled as a large spin). The efficient classical simulatibility is to our direct advantage for assessing the relative performance of the different bosonic encodings in reproducing the time-dynamics of the Dicke model as we do in this section. We expect our simulation results to carry over to the harder-to-simulate disordered model. 

Let us discuss a few known physics features of the Dicke model. In the thermodynamic limit $N_a \to \infty$, when increasing $g$ beyond a critical value $g_c = \sqrt{\omega\omega_0}/2$, one drives a phase transition for the ground state. For small $g< g_c$, one is in a normal phase with vanishing boson number density $\lim_{N_a\rightarrow \infty}\frac{\langle a^{\dagger}a\rangle}{N_a}\rightarrow 0$, while for $g > g_c$ one has a superradiant phase with macroscopic photon occupation~\cite{Emary_2003}. Indeed, if one takes the classical limit of the Dicke model for $N_a \rightarrow \infty$, the large spin represented by the atom-qubits can be modeled as an oscillator and the model becomes that of two coupled oscillators, which can show chaotic behavior in the superradiant phase. Energy level-spacing statistics which crosses over from a Poisson distribution to a Wigner-Dyson distribution has been numerically explored in e.g.~\cite{TB:OTOC,SFF:Dicke}.

For any finite $N_a$, when the coupling $g$ is small compared to the frequency gap $|\omega-\omega_0|$, one often removes counter-rotating terms (like $a^{\dagger} \sigma_i^+$, aka rotating-wave approximation) as the transfer of energy between atoms and bosonic mode
in this off-resonant regime is not efficient. With this approximation, the Dicke model is called the Tavis-Cummings model. For the Tavis-Cummings model the total number of excitations is conserved over time, and the model is integrable. Without such a rotating-wave approximation, there is no conservation of the total number of excitations and hence the number of excitations in the bosonic mode can grow large, in particular in the superradiant regime \footnote{One can derive an upperbound $\langle a^{\dagger}a \rangle(t)=O(N_a)$ for any $t$ based on energy conservation, i.e. one starts with a vacuum state and a state of maximal energy for the atom-qubits, $E_0=\frac{\omega}{2}+N_a \frac{\omega_0}{2}$. Using the Cauchy-Schwartz inequality one can bound the contribution from $\langle J_x \otimes (a+a^{\dagger})$ by $O(N_a \sqrt{\langle a^{\dagger} a\rangle(t)}))$. In the superradiant regime it is thus only the ratio $\langle a^{\dagger}a \rangle(t)/N_a$ that can grow due to large coupling $g$.}.

\begin{figure}[htb]
\centering
\includegraphics[width=0.48\textwidth,height=0.32\textheight,keepaspectratio]{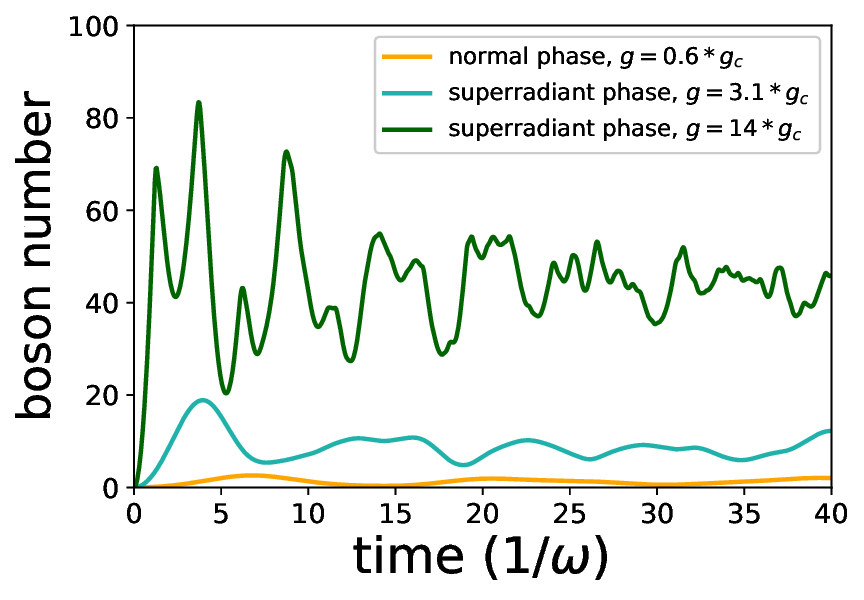}
\caption{Boson number $\langle a^{\dagger}a \rangle(t)$ under the Dicke Hamiltonian $H$ in Eq.~\eqref{eq:H_dicke}, starting in the vacuum state for the bosonic mode and $\ket{1}^{\otimes N_a}$ for the atom-qubits (with $N_a = 3$). We take $\omega = \omega_0$ here. }
\label{fig:bosonumber}
\end{figure}

In Fig.~\ref{fig:bosonumber} we demonstrate this by plotting $\langle a^{\dagger}a\rangle(t)$ for $N_a=3$ starting with the vacuum state for the bosonic mode and the three atom-qubits each in the excited state $\ket{1}$. 
The number of excitations in the bosonic mode can be far beyond the initial three excitations. It requires $N_{\rm cutoff}$ in this regime to be large, leading to a sizeable cost in space-time overhead for encoded quantum simulation, see e.g.~Fig.~\ref{fig:cnotcount_bosonumber_superradiant}.

\subsection{Circuit decompositions for quantum simulation} \label{sec:Hdicke_trotter}
\label{sec:circuitimplementation}

This section discusses the circuit decompositions for simulating time evolution of the Dicke model via the three encodings. Using a classical simulator, we numerically implement the ideal dynamics and the circuit-level dynamics to assess a level of accuracy, and we do a costing of the circuit-level dynamics. Here we provide an overview of how we obtain the circuit-level dynamics and the costing thereof. The circuit-level decomposition of the encoded dynamics of the Dicke model will use additional ancilla qubits. In all encodings we use at most $N_b$ extra ancilla qubits. 

First, in all encodings we approximate the time evolution by the encoded Dicke Hamiltonian using a simple first-order Trotterization, i.e. 
\begin{equation}
    e^{-iH t} \: \to \: \big(e^{-iH_{b,E}t/n^D_T}\;e^{-iH_{a,E}t/n^D_T}\;e^{-iH_{ab,E}t/n^D_T}\big)^{n^D_T},\label{eq:Hdicke_trotter}
\end{equation}
with $n^D_T$ Trotter layers, where the encoded Hamiltonians $H_{b,E}, H_{a,E}, H_{ab,E}$ use the descriptions in Sections \ref{sec:unifock}, \ref{sec:bindicke} and \ref{sec:binquad}.
For a single Trotter layer, $e^{-iH_{a,E}t/n^D_T}$ is a product of single-qubit rotations in all three encodings. The free-boson time evolution $e^{-iH_{b,E}t/n_T^D}$ was described in Section \ref{sec:freeH} for the three encodings, requiring either only single-qubit rotations (unary Fock and Dicke) or single-qubit rotations and QFTs (binary-x). As for the interaction, we can write 
\begin{align}
e^{-iH_{ab,E}t/n_T^D}=\Pi_{j=1}^{N_a}e^{-i \frac{gt}{n_T^D \sqrt{N_a}}X_j \otimes (\sqrt{2} x_E)},
\label{eq:cdis}
\end{align}
as all terms commute, hence we could execute this interaction fully sequentially via atom-qubit controlled displacements. 

\subsubsection{Edge coloring}
\label{sec:edge}
For sufficiently small $N_b$ and $N_a$, the sequential execution in Eq.~\eqref{eq:cdis} is not optimal in terms of circuit depth, even when we use the GHZ trick that we describe in the next section. Consider this for Dicke encoding. The dynamics in Eq.~\eqref{eq:cdis} requires each atom-qubit to partake in a two-qubit gate with each bosonic qubit, with all gates commuting. The depth to execute this circuit is $\max(N_a,N_b)$ as it is the number of colors in a graph edge coloring problem of a bipartite graph which is given by the maximum degree of any vertex in the graph (Kőnig’s line coloring theorem). For binary-x encoding, since $x_E$ is a sum over single-qubit $Z$ gates, the situation is identical and the depth is $\max(N_a,N_b)$. 
For unary encoding, we can first Trotterize $\exp(-i H_{ab,E}t/n_T^D)\approx \exp(-i H_{ab,E}^{e}t/n_T^D)\exp(-i H_{ab,E}^{o}t/n_T^D)$, i.e. two layers (even or odd) of nearest-neighbor two-qubit gates on the bosonic qubits controlled by all atom-qubits. For one layer, say the even one, since all controlled-displacements act the same on the bosonic qubits, hence commute, we can again view minimizing the depth as an edge coloring problem of a bipartite graph. In this case we have $N_a$ vertices on one side and $\lceil N_b/2\rceil$ vertices, so the depth is $\max(N_a, \lceil N_b/2 \rceil)$. Doing this both for the even and for the odd layer, gives depth at most $\max(N_a,N_b)$. 

For larger $N_b$, this strategy is not optimal in minimizing depth if we additionally allow for $N_b$ ancilla qubits. We then first use Eq.~\eqref{eq:cdis} to sequentially couple each atom-qubit to the bosonic qubits. Next, we consider the controlled-displacement for a single atom-qubit using the displacement representations in Section \ref{sec:displacement_encoding}. 

\subsubsection{Atom-qubit controlled displacement: minimizing depth}
\label{sec:mindepth}
The atom-qubit controlled-displacement of the form $\exp(-i \theta X_j \otimes x_E)$ in Eq.~\eqref{eq:cdis}, requires an `overuse' of atom-qubit $j$ as the encoded displacement on the bosonic qubits acts on all those bosonic qubits, see Section \ref{sec:displacement_encoding}.
Since the interaction only uses one type of Pauli on the atom-qubit, ---in this case Pauli $X$--, we can encode the atom-qubit temporarily in a phase-flip repetition code such that the logical Pauli $X$ is any Pauli $X$ on the code qubits (see e.g. previous uses of this ``GHZ-trick" in~\cite{StroeksClinton}). For all encodings, we use a repetition code with $N_b$ qubits. The mode of operation is then as follows. Each atom-qubit $j$ is in principle represented by a single qubit, and only prior to it being used to realize $\exp(-i \theta X_j \otimes x_E)$, it gets `folded-out' into $N_b$ qubits using a CNOT encoding circuit of depth $\lceil \log(N_b)\rceil+1$ and $N_b-1$ CNOT gates.
After the interaction, we undo this decoding, so again we represent each atom-qubit by a single qubit. So only one atom-qubit is expanded into $N_b$ qubits at a time.

In the Dicke encoding, the encoded displacement consists of single-qubit gates, hence controlled-displacement done in this manner leads directly to a circuit with only 2-qubit gates, from which the CNOT count readily follows. 

For unary Fock encoding, for each atom-qubit we use a brick-work like circuit with even and odd terms as $t/n_T^D$ is small, using $H_{ab,E}=H_{ab,E}^{e}+H_{ab,E}^{o}$. Thus the encoded displacement consists of two layers (even or odd) of nearest-neighbor two-qubit gates which are now controlled by an atom-qubit. Since these layers are consecutive, we can encode the controlling atom-qubit once before the first layer and decode only after the second, saving some CNOT gates. The resulting 3-qubit gates are of the form $\exp(-i \theta P)$ for some $\theta$, where $P$ is a 3-qubit Pauli can be further broken down into CNOT gates in a standard manner.

For the binary-x encoding, the encoded displacement uses single-qubit $Z_0$ gates, the QFT and bosonic-qubit single-qubit $Z$-rotations, and each gate in this circuit is now controlled by the atom-qubit encoded in the $N_b$-qubit phase-flip repetition code. The phase factor in Eq.~\eqref{eq:dis_x} now becomes a single-qubit $Z$ rotation on the atom-qubit which should be executed prior to the encoding into the phase-flip repetition code. One can then translate the single-qubit and CNOT count of a QFT to a corresponding 2-qubit and 3-qubit gate count and decompose the 3-qubit gates into CNOTs again. 


Details of the final CNOT counts and depths of these circuit decompositions for Dicke Hamiltonian evolution and a bound on the incurred Trotter error can be found in Table~\ref{tab:encoding-comparison} in Appendix \ref{app:circuitdecompDicke}. In this Table we did not use the edge coloring decomposition described above, although this may have been a better choice for $N_a=3$ and small $N_b$ in Fig.~\ref{fig:depth_comparison} in Appendix \ref{app:supp_info} (blue lines).

\begin{figure*}[htb]
    \centering
    \subfloat[\textbf{Normal phase} $g = 0.6*g_c$, \linebreak $\epsilon_{*} = 0.3$.\label{fig:eps03.}]{
        \includegraphics[width=0.282\textwidth,height=0.32\textheight,keepaspectratio]{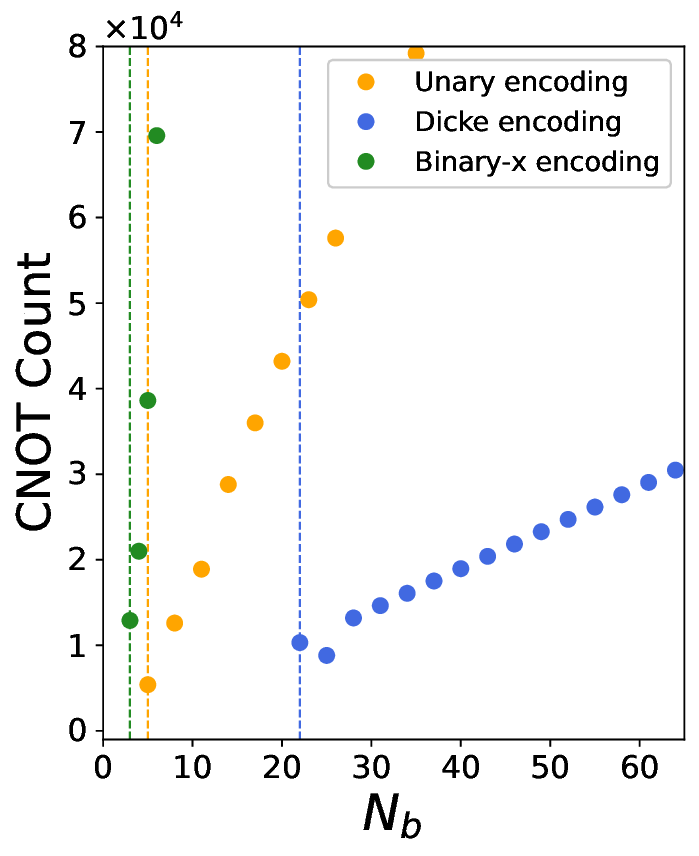}
    }
    \hfill
    \subfloat[\textbf{Normal phase} $g = 0.6*g_c$, \linebreak $\epsilon_{*} = 0.1$.\label{fig:eps01.}]{
        \includegraphics[width=0.29\textwidth,height=0.32\textheight,keepaspectratio]{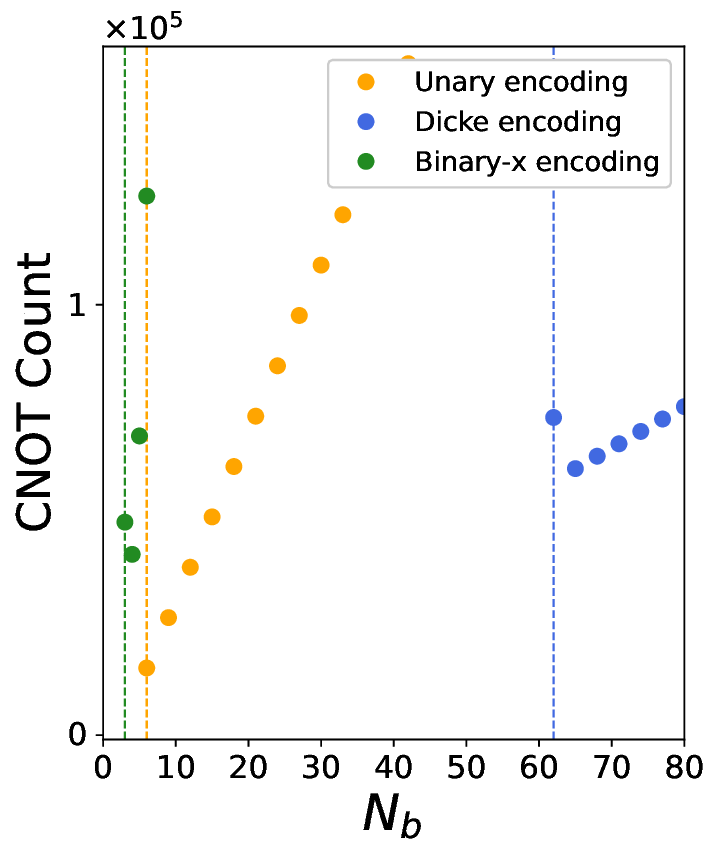}
    }
    \hfill
    \subfloat[\textbf{Super-radiant phase} $g = 3.1*g_c$, \linebreak $\epsilon_{*} = 2.2$.\label{fig:cnotcount_bosonumber_superradiant}]{
        \includegraphics[width=0.282\textwidth,height=0.31\textheight,keepaspectratio]{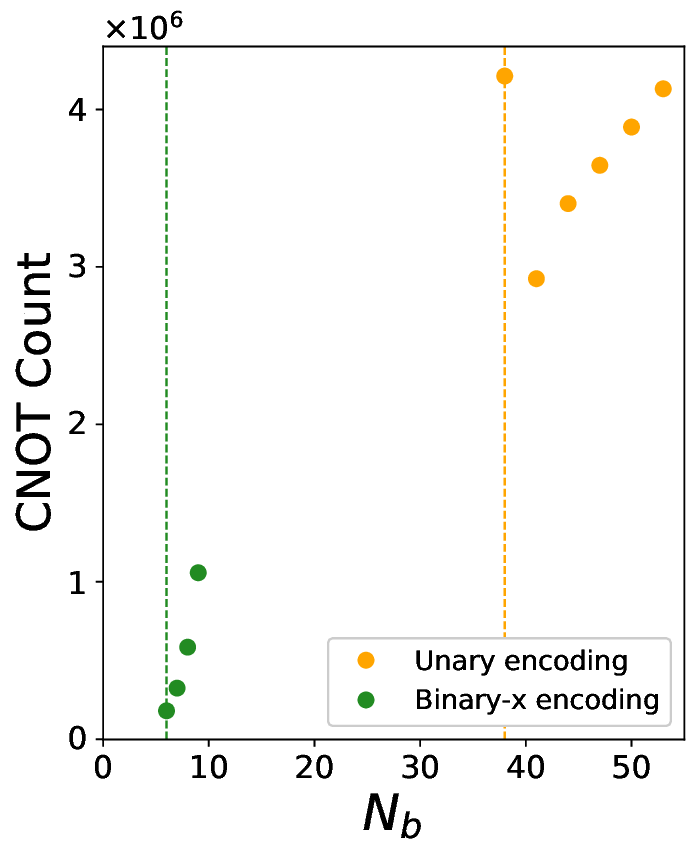}
    }
    \caption{CNOT count required to achieve an error $\epsilon_{*}$ on the average boson number of a time-evolved vacuum state. In (a,b), we take $g = 0.6*g_c$ (i.e., the normal phase), and $\epsilon_{*}=0.3$ (being $10\%$ of maximum boson number over simulated time interval) and $\epsilon_{*}=0.1$. In (c), we have $g = 3.1*g_c$ (i.e., the super-radiant phase), and $\epsilon_{*}=2.2$ (also being $10\%$ of maximum boson number over simulated time interval). Throughout, we take $N_a = 3$, and the total time interval $40/\omega$ (with $\omega = \omega_0 = 1$). The CNOT count includes the preparation of the (approximate) vacuum state 
    for the binary quadrature encoding as discussed in Appendix \ref{sec:gauss}. We pick $K = \lceil 2^{3/4*N_b} \rceil$, informed by Figure \ref{fig:AppBb} in Appendix \ref{sec:gauss}. We simulate up to $N_b\sim 10$ for the binary quadrature encoding due to memory constraints. Note that in (c), the Dicke encoding data is not shown as we simulated up to $N_b = 1000$ and even for those $N_b$'s the encoding error exceeded $\epsilon_{*}$. The vertical lines indicate the lowest $N_b$ values for which error $\epsilon_{*}$ can be achieved. }
    \label{fig:cnotcount_bosonumber}
\end{figure*}

\subsection{Numerics: comparison between encodings for Dicke model time evolution}
\label{sec:performance}

In this section, we compare the performance of the three bosonic encodings via classical simulations of the Dicke model. We implement the approximate ---encoded, Trotterized--- dynamics and compare this with ideal dynamics with respect to several output measures. We evaluate the CNOT and circuit depth cost of the approximate dynamics using the descriptions in the previous section and Table \ref{tab:encoding-comparison} in Appendix \ref{app:circuitdecompDicke}. In all simulations the initial bosonic state is the (encoded) vacuum state and the state of the atom-qubits is the excited state $\ket{1}$. 

\subsubsection{Output: expected boson number} 

In any quantum simulation the final output of the time-evolved state should be obtained efficiently and the expected boson number $\langle a^{\dagger}a\rangle(t)$ on the bosonic mode is a natural figure of merit which is efficient to compute. In the unary and Dicke encodings $a_E^{\dagger}a_E$ is diagonal in the $Z$-basis: it can be straightforwardly evaluated by measuring each qubit in the standard basis. 
In the binary-x encoding, we have to evaluate $\frac{1}{2}\langle x_E^2+p_E^2-1 \rangle$ with $x_E$ and $p_E$ in Eq.~\eqref{eq:xbin} and Eq.~\eqref{eq:pbin}. This can be realized by determining $\langle x_E^2 \rangle$ for half the runs and $\langle p_E^2 \rangle$ for the other half.
Measuring $x_E$ (and then squaring) requires measuring in the standard basis while measuring $p_E$ requires doing a QFT prior to such measurement. However, a QFT followed by a measurement can be replaced by a simpler circuit which does not use any CNOT gates, where single-qubit gates depend on the outcomes of measured qubits, see e.g. the discussion around Figure 3 in Ref.~\cite{ClassicalQFT}. Thus, also in the binary-x encoding one can measure the boson number (or, say, $p_E$) without any additional CNOT gates. 


\subsubsection{Resource estimates}
\label{sec:res_est}
We are interested in the resources required to achieve a certain error $\epsilon_{*}$ on an output $f(t)$ -- for instance, $f(t)=\langle a^{\dagger} a\rangle (t)$. For the encoded function $f_{E}(t)$ (i.e., with only an encoding error) and the encoded, Trotterized function $\tilde{f}_E(t)$, we define the following errors
\begin{align}
    &\epsilon_{E} = \max_{t}\:\lvert f(t) - f_{E}(t) \rvert, \nonumber \\ 
    &\epsilon_{\rm tot} = \max_{t}\:\lvert f(t) - \tilde{f}_E(t) \rvert.
\end{align}
We estimate the required resources as follows. First, we increase $N_b$ to the point where $\epsilon_{E}\leq \epsilon_{*}$ at some maximal time $t$. Then, for each $N_b$ above that point, we pick a minimum number of Trotter layers $n_{T}^D(t')$ s.t. $\epsilon_{\rm tot}\leq \epsilon_{*}$ at all times $t'\leq t$. As ansatz for the reduced number of Trotter layers for shorter times, we take $n_T^D(t')= \lceil n_T^D t'/t\rceil$. 

We then provide the CNOT count and circuit depth of the resulting circuits as a function of $N_b$ using Table \ref{tab:encoding-comparison} in Appendix \ref{app:circuitdecompDicke}. We consider the CNOT count because it is the usual proxy for the sensitivity of the circuits to noise. Note that the \textit{total} CNOT count -- as opposed to the CNOT count in the backward lightcone of some subset of qubits -- is the relevant quantity here, as the observable $f(t)$ we consider typically has support on the entire bosonic register. Note that we do not include the sampling error (due to having a bounded number of shots per circuit) as an explicit source of error, as we numerically observe that the sampling error is comparable across the different encodings at $\sim 10^3$ shots (in several parameter regimes), namely the sampling error on the boson number is $\sim 10^{-2}-10^{-1}$ across the encodings. 

\subsubsection{Numerical results}
\begin{figure}[htb]
\centering
\includegraphics[width=0.48\textwidth,height=0.22\textheight,keepaspectratio]{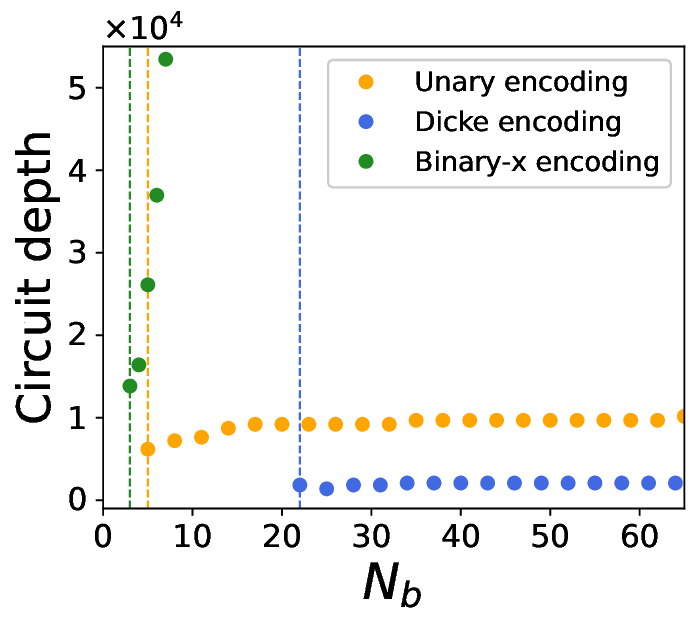}
\caption{Circuit depth required to achieve an error $\epsilon_{*} = 0.3$ ($10\%$ of maximum boson number over simulated time interval) on the average boson number of a time-evolved vacuum state. We take $g = 0.6*g_c$, $N_a = 3$, and the maximal time is $40/\omega$ (with $\omega = \omega_0 = 1$ here). The circuit depth includes the preparation of the (approximate) vacuum state 
for the binary quadrature encoding as discussed in Appendix \ref{sec:gauss}. We pick $K = \lceil 2^{3/4*N_b} \rceil$, informed by Figure \ref{fig:AppBb} in Appendix \ref{sec:gauss}. The vertical lines indicate the lowest $N_b$ values for which error $\epsilon_{*}$ can be achieved. }
\label{fig:circuitdepth_bosonumber}
\end{figure}

Figure~\ref{fig:cnotcount_bosonumber} gives the CNOT count required for estimating the average boson number within an accuracy $\epsilon_{*}$ for normal-phase and super-radiant-phase configurations for $N_a=3$
-- see Figure~\ref{fig:bosonumber} for the associated boson number dynamics. For any encoding, there is a minimum $N_b^{*}$ below which the encoding error exceeds $\epsilon_{*}$. Above this minimum $N_b^{*}$, the CNOT count increases with $N_b$ -- making the circuit at $N_b^{*}$ the optimal choice in terms of CNOT count. Simulating time dynamics in the super-radiant phase to a fixed absolute precision requires larger circuits since the larger average boson number demands a larger $N_b$ to ensure $\epsilon_{E} < \epsilon_{*}$, see also Figure \ref{fig:encoding_errors}. This effect is especially important for the Dicke encoding, for which $\epsilon_{E}$ exceeds $\epsilon_{*}$ in Figure \ref{fig:cnotcount_bosonumber_superradiant} even at $N_b = 1000$ (beyond which our numerical simulations become rather slow). It is clear from this data that the CNOT counts are comparatively low for the binary-x encoding: this also goes hand-in-hand with the binary-x encoding using the least number of qubits $N_b$, which can be viewed as the advantage of encoding information in binary and not unary. However, when we instead plot the circuit depth for the parameters of Fig.~\ref{fig:eps03.}, we observe that circuit depth is lowest for the Dicke encoding, see Fig.~\ref{fig:circuitdepth_bosonumber}. 

When only fixing the error on the boson number as we do here, one observes from Figures~\ref{fig:cnotcount_bosonumber} and~\ref{fig:circuitdepth_bosonumber} that the number of Trotter layers is approximately constant as a function of $N_b$ as soon as $\epsilon_{E} < \epsilon_{*}$. This is especially clear from the linear CNOT count dependence for the unary and Dicke encoding in Figure \ref{fig:cnotcount_bosonumber} and the approximately constant circuit depth for the unary and Dicke encoding in Figure~\ref{fig:circuitdepth_bosonumber}. On the other hand, the Trotter error column and CNOT count/depth per layer column in Table~\ref{tab:encoding-comparison} in Appendix~\ref{app:circuitdecompDicke} suggests that fixing the error with respect to $||\cdot||_E$ requires the unary and Dicke encoding CNOT count to scale superlinearly and the unary and Dicke encoding circuit depth to scale superlogarithmically in $N_b$. However, this is not observed for this error regime and the boson number output (which could certainly be a less-distinguishable error metric).


\section{Experiments on the H2-2 Quantinuum Device}
\label{sec:h22}

We have performed two experiments on Quantinuum's H2-2 56 trapped-ion qubit device~\cite{quantinuum_h22,Moses_2023}. We tested the performance of both unary and Dicke encoding in this noisy setting. In total we use at most $2N_b+N_a$ qubits on the device in the simulations.

To fix the circuits that we run, we proceed as in Section~\ref{sec:res_est}. Namely, we set a target error $\epsilon_*$ on the final observable that we measure and, given an encoding, we choose the smallest $N_b$ which allows the encoding error to be (sufficiently) less than $\epsilon_*$, using a rescaling ansatz of the number of Trotter layers. So, also for the displacements, we pick a number of Trotter layers $n_T(\alpha')= \lceil n_T \alpha'/\alpha\rceil$ sufficiently large so that the total error on the output observable for any $|\alpha'| \leq |\alpha_{\rm max}|$ is below the targeted error $\epsilon_*$ (without taking into account any sampling error).  

The resulting circuits then still need to be compiled to the hardware device. For this, we have used pytket~\cite{Pytket,Sivarajah_2021} and qnexus~\cite{quantinuum_nexus}: when compiling the circuits with pytket we use optimisation level 3. For more information on the specific device backend, optimisation level, and software, see Appendix~\ref{app:supp_info}. In particular, reasons for discrepancies between our theoretical CNOT count and depth estimates, and those of the compiled circuits, are briefly discussed in Appendix \ref{app:opti}.
For every circuit (including measurement) to be ran on the device we use $M_{\rm samp}=500$ or $M_{\rm samp}=1000$ shots per circuit for both encodings and both experiments.  


\subsection{Displacement experiments} \label{sec:experiment_displacement}

One experiment is the realization of the displacement $D(\alpha)$ in Eq.~\eqref{eq:displacement} with a purely imaginary $\alpha$ on the vacuum state, which has $\langle p\rangle =\sqrt{2}|\alpha|$ in the ideal case. For unary Fock encoding, we Trotterise the displacement operator as in Eq.~\eqref{eq:displacement_unary_trotter}. We measure the expectation of $p_E$ which, for unary encoding, is given by:
\begin{equation}
p_E = \frac{1}{2\sqrt{2}} \sum_{j=0}^{N_b-2} \sqrt{j+1} \left(X_j Y_{j+1} -Y_j X_{j+1}\right).
\end{equation}
The terms in this sum can be split, $p_E=p_{e,E}+p_{o,E}$ where $p_{e,E}$ includes all pairs of indices $(j,j+1)$ for $j$ even, and $p_{o,E}$ for $j$ odd. In order to measure, say, $p_{e,E}$ we first do a two-qubit Clifford rotation on all pairs of qubits so that $X_jY_{j+1}$ and $Y_j X_{j+1}$ get mapped onto single-qubit Pauli $Z_j$ and $Z_{j+1}$. Thus, to measure the expectation value for $p_E$, we estimate $p_{e,E}$ for half the circuits and $p_{o,E}$ for the other half. We ran each each circuit for a given $\alpha$ for $M_{\rm samp}=500$ shots, computed $p_{e,E}$ or $p_{o,E}$ for each individual shot, and then took the mean and added $p_{e,E}$ and $p_{o,E}$ in the end.  


For Dicke encoding, we use the depth-1 circuit given by Eq.~\eqref{eq:displacement_dicke} and measure the expectation value of
\begin{equation}
p_E=\frac{1}{2 \sqrt{2N_b}} \sum_{j=0}^{N_b-1} Y_j,
\end{equation}
by measuring each qubit in the $Y$ basis, again using $M_{\rm samp}=500$ shots in total.

\begin{figure}[htbp]
    \centering
    \subfloat[\label{fig:trotter_comparison}]{
        \includegraphics[width=0.7\linewidth]{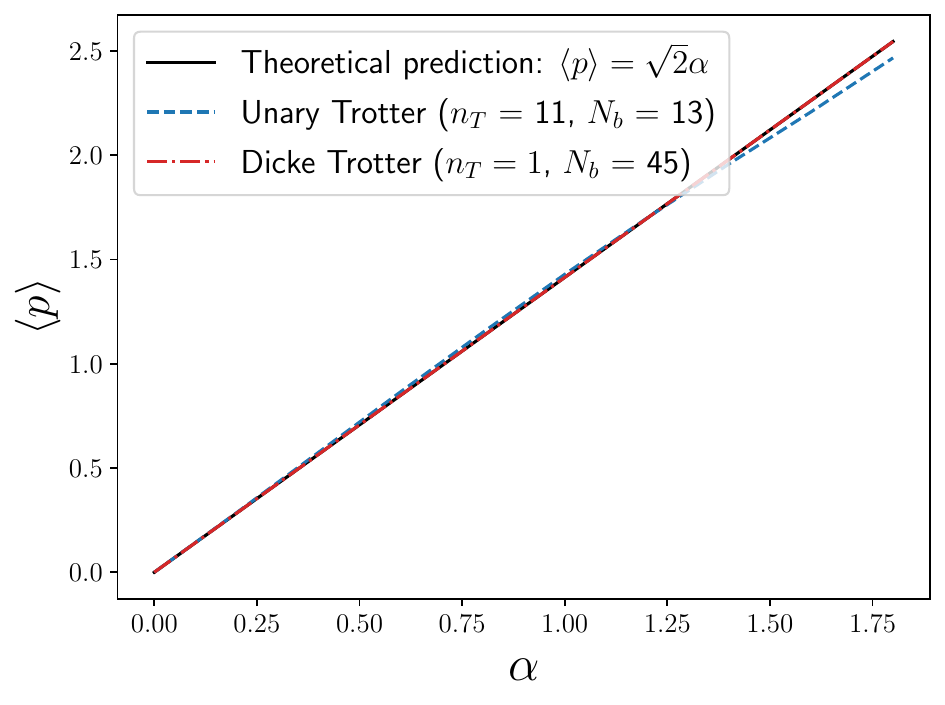}
       }
       \vspace{0.2cm}
    \subfloat[\label{fig:unary_and_dicke}]{
        \includegraphics[width=0.7\linewidth]{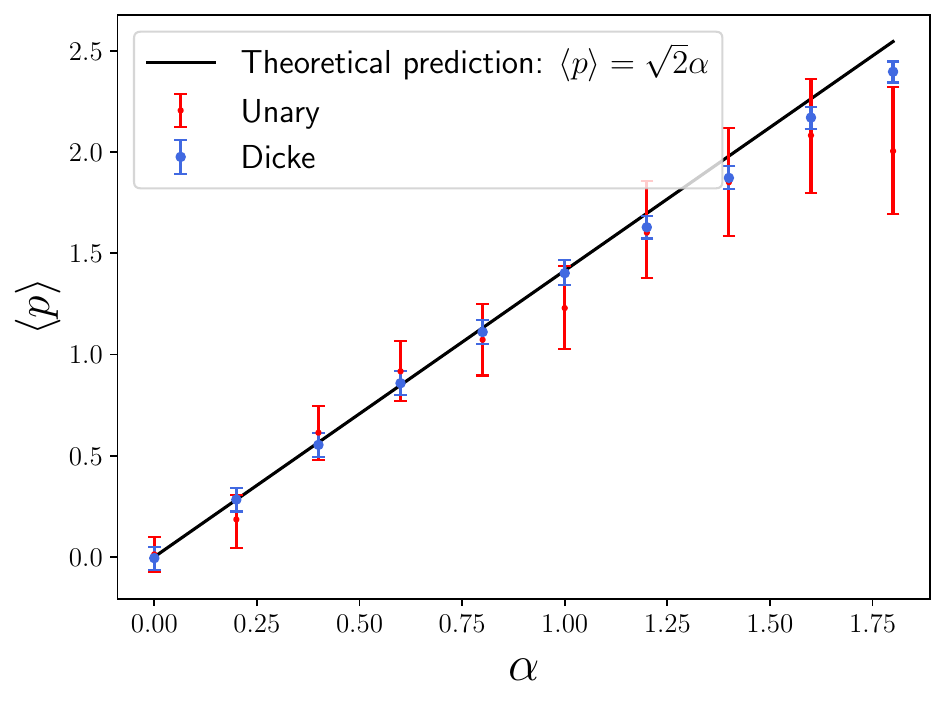}
        }
    \vspace{0.2cm}
    \caption{Simulation of a displacement in the $p$-direction on Quantinuum's H2-2 using unary Fock and Dicke encoding. (a) Exact evolution versus noise-free, encoded, Trotterised evolution using $N_b=13, n_T=11$ for unary Fock encoding and $N_b=45$ for Dicke encoding. $N_b$ and $n_T$ were chosen such that the maximum absolute error on $\langle p\rangle$ is less than $\epsilon_*=0.1$. (b) Quantum simulation of the compiled circuits for these encoded displacements on the Quantinuum device. The error bars are 95\% confidence intervals estimated via bootstrapping with 1000 resamples.}
    \label{fig:unary_dicke_comparison_displacement}
\end{figure}

The resulting data for the simulation of a displacement in $p$-direction is plotted in Fig.~\ref{fig:unary_and_dicke} where we have chosen an $\epsilon_*=0.1$ error on the estimate for $\langle p_E\rangle$. The error bars for both Dicke and unary encoding were estimated via bootstrapping~\cite{bootstrap} with 1000 resamples. We estimate the confidence interval of $p_{e,E}$ and $p_{o,E}$ from their individual confidence intervals using the union bound. For a more detailed discussion of the bootstrapping process, see Appendix~\ref{sec:supp_info_error}. 

The resulting depth of the compiled circuit for unary encoding for the largest $\alpha$ was 86 and its two-qubit gate count was 210 (which includes the two-qubit rotations for the final measurement). Our theoretically expected depth used in our classical simulations and based on Table~\ref{tab:displacement-encoding-comparison} is $18\cdot n_T=198$ for unary encoding and its CNOT gate count is $4 (N_b-1) \cdot n_T=528$ (excluding the final two-qubit measurement). 

Not surprisingly, Dicke encoding, which only uses single-qubit gates, tracks the noise-free evolution exactly. Since this is a circuit with only single-qubit gates for any $\alpha$, the noisy evolution should also not get worse with increasing $\alpha$ (although the observable that we estimate grows with $\alpha$), but we do seem to see a slight under-performance at large values of $\alpha$ in Fig.~\ref{fig:unary_dicke_comparison_displacement}.
For the unary Fock encoding, the noisy performance shows larger deviations due to the circuit being larger and increasing in size with $\alpha$ and $n_T$. One wonders whether choosing a lower number of Trotter layers, hence tolerating a larger error $\epsilon_*$ in the noise-free case, would have led to overall better performance in the presence of noise.







\vspace{-0.2cm}
\subsection{Time evolution of the Dicke model}\label{sec:experiment_timeevo}
\vspace{-0.3cm}

We have implemented the time evolution $\exp(-iH_Et)$ for unary and Dicke encoding using the Trotterisation and circuit decompositions discussed in Section~\ref{sec:Hdicke_trotter}. Similar, as in our classical simulations we take $N_a=3$. Again, we set a value for the error on the output, i.e. $\langle a^{\dagger}a \rangle(t)$ (boson number), and we choose $\epsilon_* = 0.025$: we pick the smallest $N_b$ and the ansatz $n_T^D(t')=\lceil n_T^D t'/t \rceil $ such that the encoded Trotterized evolution for all chosen times $t'\leq t$ produces $\langle a^{\dagger}a \rangle(t)$ with an error below $\epsilon_*$. We do not implement the GHZ trick discussed in Section~\ref{sec:mindepth} for the unary encoding as $N_b=3$ which is very small. For the Dicke encoding we do use the GHZ trick with $N_b=5$.

\begin{figure}[htbp]
    \centering
     \subfloat[\label{fig:unary_dicke_comparison_time}]{
    \includegraphics[width=0.7\linewidth]{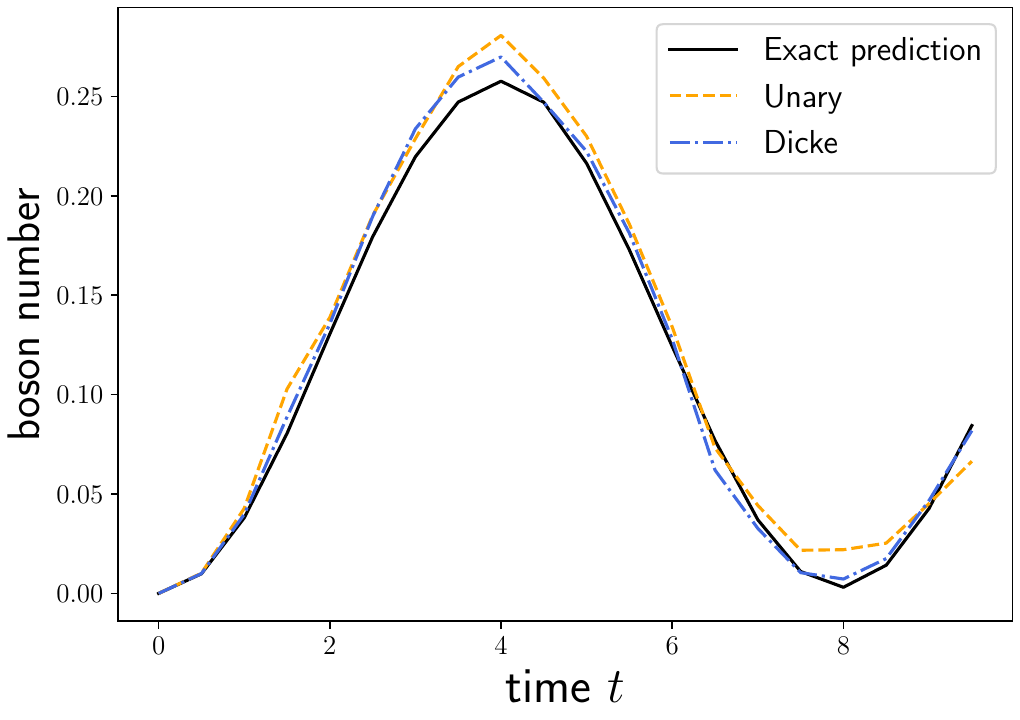}}
    \vspace{0.2cm}
    \subfloat[\label{fig:unary_and_dicke_time}]{
        \includegraphics[width=0.7\linewidth]{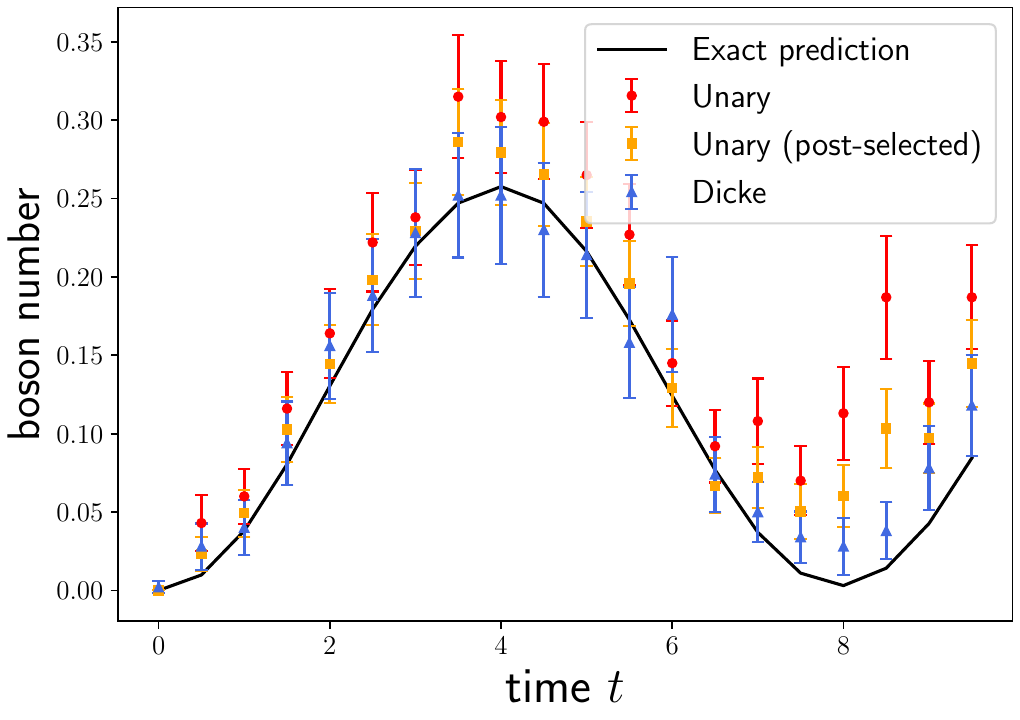}}
    \caption{Simulation of time evolution induced by the Dicke Hamiltonian for $N_a=3$, $g=0.2$ ($=0.3*g_c$), $\omega=1$, $\omega_0=1.8$. (a) Exact evolution versus noise-free Trotterised encoded evolution using $N_b=3$ and $n_T^D=6$ for unary Fock encoding, and $N_b=5$, $n_T^D=6$ for Dicke encoding. (b) Quantum simulation of the compiled circuits on the Quantinuum H2-2 device. The error bars are 95\% confidence intervals estimated via bootstrapping with 1000 resamples.}
\end{figure}

The measured expectation values $\langle a^{\dagger}a \rangle(t)$ for unary and Dicke encoding over time are shown in Fig.~\ref{fig:unary_and_dicke_time}. Figure~\ref{fig:unary_dicke_comparison_time} shows the theoretically predicted time evolution that we obtain by numerically solving the Schr\"odinger equation with QuTip~\cite{qutip5} and the evolution for the error-free encoded Trotterisation. Note that our choice for parameters puts us in the very low-boson number off-resonant regime where quantum simulation is not expected to be hard.
The post-selected data for the unary encoding in Fig.~\ref{fig:unary_and_dicke_time} are obtained by tossing out runs where one measures a bit string which has more than a single 1 somewhere (so out of the encoded space), and such post-selection clearly improves the data. Note that bit-flip noise can have a large detrimental effect in the unary encoding due to the $j$ prefactor in Eq.~\eqref{eq:num_operator_unary}. Such sensitivity is not present for the boson number in Dicke encoding; however it is unclear how one can post-select on being in the permutation-symmetric `Dicke' subspace without using additional resources.

We see that unary encoding, with or without post-selection, deviates from the theoretically-predicted curve for large times, suggesting slightly more advantageous behavior of Dicke encoding in the presence of noise, presumably due to the lower two-qubit gate count and depth. To be precise, the resulting depth after compilation for unary and Dicke encoding at the last time step are 111 and 56 respectively (compared to 582 and 204 as theoretically predicted, see Table~\ref{tab:encoding-comparison}). The two-qubit gate count at the last time step are 93 and 100 respectively (compared to the predicted CNOT gate of 180 and 324). 
For the same parameter value and error $\epsilon_*$ we note that the binary-x encoding would have led to $N_b=4$ and $n_T^D=50$ leading to a theoretically-predicted CNOT count of $3700$ and, coincidentally, a depth of $3700$ (when including the GHZ trick). Note that we aim at a lower value of $\epsilon_*$ here than in the plots in Section \ref{sec:performance}. Due to having only limited (promotional) access to the trapped-ion qubit device, we did not implement these experiments for the binary-x encoding.




\section{Discussion}
\label{sec:discuss}

In the paragraphs below we discuss a few open problems and ideas. 

\subsection{Other benchmarks}
In this paper we have chosen simple metrics, namely realizing displacements and implementing time evolution of the Dicke model to compare bosonic encodings. One could explore the performance of such encodings further by asking whether they can reproduce interesting `physics' such as the level statistics, spectral form factors or OTOCs of a disordered Dicke model, in the limit of large $N_a$~\cite{SFF:Dicke,TB:OTOC}. However, first, level statistics depends on the difference between adjacent energy eigenvalues of the Hamiltonian $H$, which is not a simple function of $H$ itself (for functions of $H$ one could employ block encoding techniques). It is also not clear to us that the limited $1/{\rm poly}(N_b)$ accuracy in eigenvalue quantum algorithms would suffice to distinguish Poissonian versus non-Poissonian behavior. Similarly, it is unclear whether efficient quantum algorithms have enough spectral resolution to predict the behavior of a (connected) spectral form factor \cite{ConnectedSFF,Martinez_Azcona_2025} $\frac{1}{2^{2n}}\Big(\bigl\langle\bigl\lvert{\rm Tr}(e^{i H t})\bigr\rvert^2\bigr\rangle - \bigl\lvert\bigl\langle{\rm Tr}(e^{i H t})\bigr\rangle \bigr\rvert^2\Big)$.
If we wish to analyze which bosonic encodings could best reproduce the `dip, ramp, plateau' features of a spectral form factor, ideally we perform a classical simulation using genuinely disordered Hamiltonians: however, the classical simulation will become rapidly infeasible for large $N_a$. Thus, classical simulations of the spectral form factor have used temporal averaging instead, see e.g. \cite{TB:OTOC}. For level statistics, one can use techniques as employed in \cite{Berke2023}. But how do we know much error we incur due to such alternative averaging, and how do we distinguish this from encoding errors? Due to these issues, we have refrained from exploring this, although experimental analog simulations consider precisely this \cite{AnalogSimulationDicke}. 

In this paper we focused on a single bosonic mode and assumed that its full quantum state was relevant in the dynamics. In contrast, many physics models use multiple bosonic modes (`a bath') ---often modeled as an open system, and often the bosons are at low-frequencies as compared to temperature---, leading to approximations where the bosonic modes are integrated out (think: phonons in BCS theory). It is an open question how one can use quantum resources efficiently to go beyond such effective dynamics.

\subsection{Use of rotating or interaction frame}
\label{sec:rot_frame}

It is common to physically implement or simulate the time evolution induced by a Hamiltonian in a rotating or interaction frame where one `co-moves' with the free uncoupled evolution of both atoms and boson(s): this picture suffices when final measurements commute with the free evolution. For example, for the Dicke model Hamiltonian in Eq.~\eqref{eq:H_dicke}, one can rotate away the free bosonic and qubit evolutions due to $H_a$ and $H_b$, and consider the simulation of a ---now time-dependent--- interaction term in such rotating frame. This interaction (with time-dependencies $e^{\pm i(\omega\pm \omega_0)t}$) will need to be Trotterized and the fact that displacements at different times are no longer commuting (so we can no longer apply the atom-qubit controlled displacements sequentially as in Eq.~\eqref{eq:cdis}) does not obviously make the simulation less costly in any of the encodings. Hence we did not pursue this approach in this paper.

\subsection{Different encodings}
Beyond the bosonic encodings discussed here and in the literature, one can imagine yet different ways to parcel up phase space: basically any method that is used to digitize and simulate bosonic modes on a classical computer (as in the QuTip, BosonQiskit or Strawberry Fields packages) can be envisioned. For example, one can use a basis of coherent states $\ket{\alpha}$, where $\alpha$ is discretized and represented by two integers, i.e. $\alpha=\alpha_{l,k}=\sqrt{\delta}(l +i k)$. For $\delta=\pi$ this basis has been shown to be a complete (but non-orthogonal) basis for the entire phase space, forming a `von Neumann lattice', see e.g. references in~\cite{Weigand2020PhD}. For larger $\delta >\pi$, the states become orthogonal, hence displacements can be represented as simple additions (times a phase). However, the accuracy of simulation is then limited by the choice of $\delta$ and phase-space rotations generated by $a^{\dagger} a$ seem awkward. One can also discretize the Wigner function representation $W_{m,m'}(x,p,t)$ of a state $\bra{m} \rho \ket{m'}$ where $m,m'$ are atom-qubit basis states and $\rho$ is the full boson-atom density matrix. One can try to represent the dynamics of this matrix, i.e. ${\bf W}(x,p,t)=e^{\mathcal{L}t}({\bf W}(x,p,0)$ plus a final measurement on a quantum computer (implementing its dynamics directly or using block-encoding). It is not clear to us that these encodings have direct advantages. 

\subsection{Logical bosons?}
\label{sec:discuss-logical}

Let's discuss some known and new ideas on how to encode logical bosons. Once `all things boson' are represented in terms of qubits, as we have done in this paper, we can apply the framework of qubit stabilizer codes and be done. So here we discuss instead having either physical qudits or physical bosons at our disposal: how can this be exploited to reduce space-time costs.

\subsubsection{Using physical qudits of dimension $d=2^m$}
We imagine that we have physical qudits of dimension $d=2^m$, and use the binary quadrature encoding which heavily uses the QFT. If the QFT for a qudit stabilizer code is transversal, then this would be advantageous in terms of space-time overhead. One can certainly take a qudit color code with {\em prime} dimension $d$ \cite{watson-qudit, moussa}, such as the qudit Steane code to enact a qudit-Hadamard, aka Quantum Fourier Transform gate, transversally. A concrete example for dimension $d=2^m$ is, say, the Steane code $[[7,1,3]]_{d=2^m}$.
Let's use the notation $\mathcal{X}$ and $\mathcal{Z}$ for the generalized qudit Pauli operators of dimension $d$ with $\mathcal{X} \mathcal{Z}=e^{-2\pi i/d}\mathcal{Z}\mathcal{X}$, $\mathcal{X}^{d}=\mathcal{Z}^{d}=I$. Let $F\equiv {\rm QFT}$ so $F \mathcal{X} F^{\dagger}=\mathcal{Z},F \mathcal{Z}F^{\dagger}=\mathcal{X}^{-1}$. The stabilizers of this qudit Steane code are
\begin{align}
    S_1^{\mathcal{X}}=\mathcal{X}_1 \mathcal{X}_2 \mathcal{X}_3 \mathcal{X}_4,S_1^{\mathcal{Z}}=\mathcal{Z}_1 \mathcal{Z}_2^{-1} \mathcal{Z}_3^{-1} \mathcal{Z}_4, \notag \\
    S_2^{\mathcal{X}}=\mathcal{X}_1 \mathcal{X}_2 \mathcal{X}_5 \mathcal{X}_6, S_2^{\mathcal{Z}}=\mathcal{Z}_1 \mathcal{Z}_2^{-1} \mathcal{Z}_5^{-1} \mathcal{Z}_6,\notag \\
    S_3^{\mathcal{X}}=\mathcal{X}_1 \mathcal{X}_3 \mathcal{X}_5  \mathcal{X}_7, S_3^{\mathcal{Z}}=\mathcal{Z}_1 \mathcal{Z}_3^{-1} \mathcal{Z}_5^{-1}  \mathcal{Z}_7.
\end{align}
 The logical operators can be chosen as $\overline{\mathcal{X}}=\mathcal{X}^{\otimes 7}$ and 
 $\overline{\mathcal{Z}}=\mathcal{Z}_1 \mathcal{Z}_2^{-1}\mathcal{Z}_3^{-1}\mathcal{Z}_4 \mathcal{Z}_5^{-1}\mathcal{Z}_6 \mathcal{Z}_7$. We have 
\begin{align}
 \overline{F}=F_1 F_2^{\dagger} F_3^{\dagger} F_4 F_5^{\dagger} F_6 F_7.
 \end{align}
 Note that the qubit Pauli on the least significant bit $Z_0=\mathcal{Z}^{2^m-1}$ and thus its logical version
$\overline{Z_0}$ is also transversal. 
Hence, the transformation from the logical $\overline{x}_E$ to $\overline{p}_E$ in Eq.~\eqref{eq:pbin} which uses the QFT and $Z_0$, can be executed transversally, in parallel on each qudit. We can also breakdown the logical rotation gates $\exp(- i\theta \overline{x_E})$ and $\exp(-i \theta \overline{x_E}^2)$ by using that 
$\overline{x_E}= \sum_{p=0}^{2^m-1} \alpha_p \overline{\mathcal{Z}}^p$ for some coefficients $\{\alpha_p\}$. Thus the cost of simulating these $\overline{x}_E$ interactions depends on the cost of simulation the non-Clifford $\mathcal{Z}$ rotations for the Steane $[[7,1,3]]_{d=2^m}$ code, and this cost can be expected to grow with $2^m$.


\subsubsection{Using physical bosonic modes}
Previous work has shown that the direct, unencoded, use of bosonic modes can lead to large savings as compared to using multiple qubits to represent a bosonic mode \cite{crane2024+, liuhybrid2026}. However, once we move into the error-corrected regime and use physical qubits to represent logical qubits, those logical qubits with high-weight logical operators may be ill-matched to interacting with such physical bosonic modes. Firstly, the noise on the physical bosonic modes may be much higher than that of the logical qubits. Secondly, one physical bosonic mode ---a superconducting resonator say---, will have to interact and thus be coupled to many physical qubits to realize a logical action.

 It would thus be advantageous to use multiple physical bosonic modes to encode logical bosonic modes. No-go theorems exist for Gaussian bosonic states when using \href{https://errorcorrectionzoo.org/c/analog_stabilizer}{Analog stabilizer codes} 
 (also called linear oscilator codes~\cite{VAHWPT}). That is, using Gaussian states, an asymmetric analog stabilizer toric code can suppress the displacement noise in one quadrature, depending on code size, but such logical noise squeezing leads to a gain in noise in the other quadrature. A proposal in Ref.~\cite{NGJ} has shown that the use of non-Gaussian GKP states can help in suppressing the noise for both quadratures. However, for increasing code size $n$: (1) the construction seems to require an energy cost which increases exponentially in $n$ (due to squeezing) while suppressing the noise variance exponentially (compare: qubit codes for which the energy cost is linear in code size $n$), (2) noise on GKP input states can get amplified due to squeezing. See also~\cite{HK:nogo, Wu2023}. A different option pursued in~\cite{Xu_2025, XYA, poirson2026} is to translate the framework of homological rotor codes of~\cite{VCT:torsion} to bosons, replacing a code constraint on the angular momenta $\ell \in \mathbb{Z}$ of sets of rotors to a constraint on the boson number of sets of bosons, see \href{https://errorcorrectionzoo.org/c/homological_number-phase}{Homological number-phase code}. 
 While the logical number operator is simple for these codes, a logical creation operator has an awkward expression, not lending itself to easy implementation on the constituent bosonic modes of the code.

\section{AI statement and code $\&$ data availability}
\vspace{-0.2cm}

We have used various AI models for small, easy-to-verify, mathematical questions and the conversion of Jupyter notebooks to Python code for the plots and data evaluation. Maybe the best current AI models could have done all the numerics and some of our analysis and interpretation. In any case, we enjoyed doing it ourselves. 

The code for the experiment and the resulting data is available in \url{https://github.com/JelenaMack/Bosonic_Simulation}. 
All code written to run the experiment was human written. Claude Opus 4.8 was used to debug code.

The code used for classical simulations is available in \url{https://github.com/MStroeks/Bosonic_Simulation} and was human written. All classical simulations in this work can be run on a standard laptop within reasonable time, i.e., each individual plot can be produced on the scale of a few hours. For the simulations that we carried out, we found that the main bottleneck is lack of memory, rather than runtime. Note that for the unary Fock and Dicke encodings, we could go up to $N_b \sim 1000$, and for the binary quadrature encoding up to $N_b = 10$. 

\section{Acknowledgements}
This work was supported by the Netherlands Organisation for Scientific Research (NWO/OCW), as part of Quantum Limits (project number summit.1.1016).  J.M. acknowledges funding from the Dutch Research Council (NWO) through the Quantum Software Consortium (NWO Zwaartekracht Grant No.~024.003.037). Research of Y.H. at Perimeter Institute is supported in part by the Government of Canada through the Department of Innovation, Science and Economic Development and by the Province of Ontario through the Ministry of Colleges and Universities. Y.H. is also supported by the Applied Quantum Computing Challenge Program at the National Research Council of Canada. We thank Quantinuum for free promotional access to the H2-2 device, and in particular Zachary Massa for introduction and answering questions. 
We thank M.H. Shaw for interesting discussions. We thank Eleanor Crane for some feedback on the paper. 

\subsection{Contributions} J.M. wrote the code for the Quantinuum device experiments and ran the experiments and did the data analysis. M.S. ran the classical simulations (with code set up by Y.H. and J.M.) with cost estimates. B.M.T. initiated the project, and B.M.T. and Y.H. supervised the project. B.M.T., M.S. and J.M. wrote the paper. 


%

\appendix
\onecolumngrid
\section{Trotter bound: displacement}\label{app:Trotter_derivation}

We will use the first-order Trotterization $\tilde{U}(t)$ of a Hamiltonian evolution $U(t)=\exp(-i H t)$ with $H=\sum_j H_j$ so that the Trotter error can be upper bounded as (see e.g. \cite{Dalzell_quantum_algorithms}):
\begin{equation}\label{eq:trotter_error}
\bigl \lVert U(t)-\tilde{U}(t) \bigr \rVert \leq \frac{t^2}{2n_T} \sum_{i} \bigl \lVert \sum_{j> i}[H_i,H_j] \bigr \rVert.
 \end{equation}
When we estimate the Trotter error {\em within} the codespace, we denote the norm as $||\cdot||_E$. We comment on how the Trotter error can get worse when one considers the entire space. 
 
We provide the derivation of the Trotter error $\epsilon_T=O\left(\frac{N_b|\alpha|^2}{n_T}\right)$ in a displacement for the unary Fock encoding in Table~\ref{tab:displacement-encoding-comparison} in Section~\ref{sec:displacement_encoding}, i.e. we bound
\begin{align}
    ||D_E(\alpha)-\tilde{D}_E(\alpha)||_E
\end{align}
 with $\tilde{D}_E(\alpha)=U_{E,T}^{n_T}(\alpha)$ with
\begin{equation}\label{eq:unary_displacement_even_odd_app}
U_{E,T} =\exp(-iH^e(\alpha/n_T) ) \exp(-iH^o (\alpha/n_T)) =\exp\left(-i \frac{|\alpha|}{n_T} H^e(e^{-i\phi_{\alpha}})\right)\exp\left(-i \frac{|\alpha|}{n_T} H^o(e^{i\phi_{\alpha}})\right),
\end{equation}
and $H^e(.)$ and $H^o(.)$ given in Eqs.~\eqref{eq:displacement_unary_He}-\eqref{eq:displacement_unary_Ho} and 
$\alpha=|\alpha|e^{i\phi_{\alpha}}$. Applying Eq.~(\ref{eq:trotter_error}) gives
\begin{align}
    ||D_E(\alpha)-\tilde{D}_E(\alpha)||_E\leq \frac{|\alpha|^2}{2n_T}||[H^e(z),H^o(z)]||_E,\; z=e^{i\phi_{\alpha}}.
\end{align}
with 
\begin{align}
   A\colon=[H^e(z),H^o(z)]=-\sum_{j=0}^{N_b-3}(-1)^j \sqrt{(j+1)(j+2)}\left[z^2 \sigma_j^- Z_{j+1} \sigma_{j+2}^+ -
(z^*)^2\sigma_j^+ Z_{j+1} \sigma_{j+2}^- \right].
\label{eq:defA}
\end{align}
When we evaluate the norm of $A$ with respect to the unary code space, we can use that it is an off-diagonal matrix with at most 2 entries per row (or column) and the strength of each entry $|A_{ij}|\leq |z|^2\sqrt{(N_b-1)(N_b-2)}$ which leads to 
\begin{align}
||A||_E\leq 2 \sqrt{(N_b-1)(N_b-2)},
\label{eq:boundA}
\end{align}
implying 
\begin{equation}\label{eq:displacement_unary_trotter_bounded_res}
 \bigl \lVert |D_E(\alpha)-\tilde{D}_E(\alpha) \bigr \rVert_E \leq \frac{2|\alpha|^2}{2n_T} \sqrt{(N_b-1)(N_b-2)} =O\left(\frac{N_b|\alpha|^2}{n_T}\right).
 \end{equation}
 When looking at $||A||$ on the entire space, one can simply bound the contribution from each sum over $j$ so that 
 $||A||\leq 2 (N_b-2)\sqrt{(N_b-1)(N_b-2)}=O(N_b^2)$.
 


\section{Preparing the vacuum state in the binary quadrature encoding}
\label{sec:gauss}

\begin{figure*}[b]
    \centering
    \subfloat[Infidelity at varying $N_b$.\label{fig:AppBa}]{
        \includegraphics[width=0.6\textwidth,height=0.25\textheight,keepaspectratio]{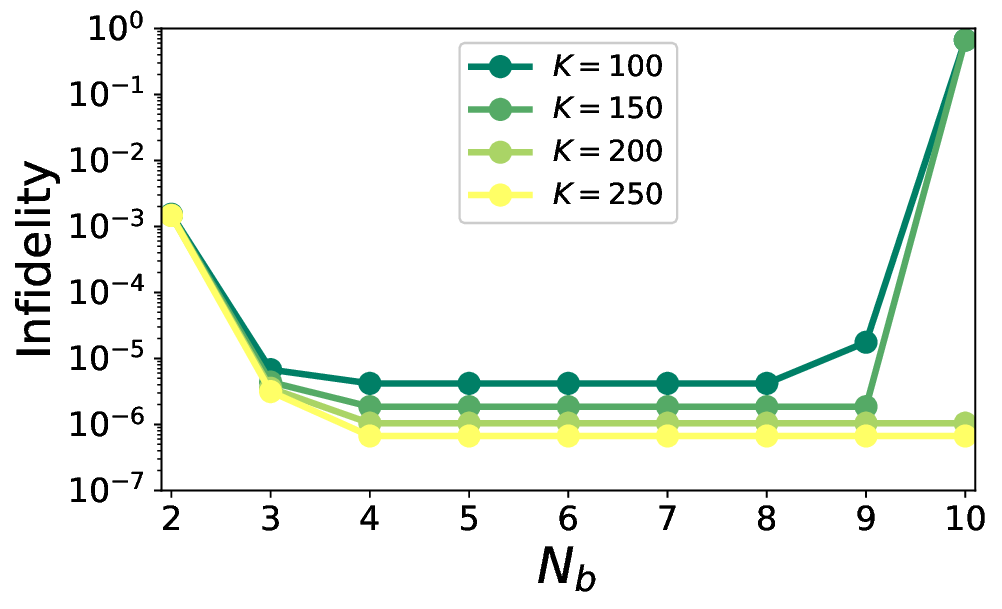}
    }
    \hfill
    \subfloat[Infidelity at fixed $N_b = 5,6,7,8$. \label{fig:AppBb}]{
        \includegraphics[width=0.4\textwidth,height=0.25\textheight,keepaspectratio]{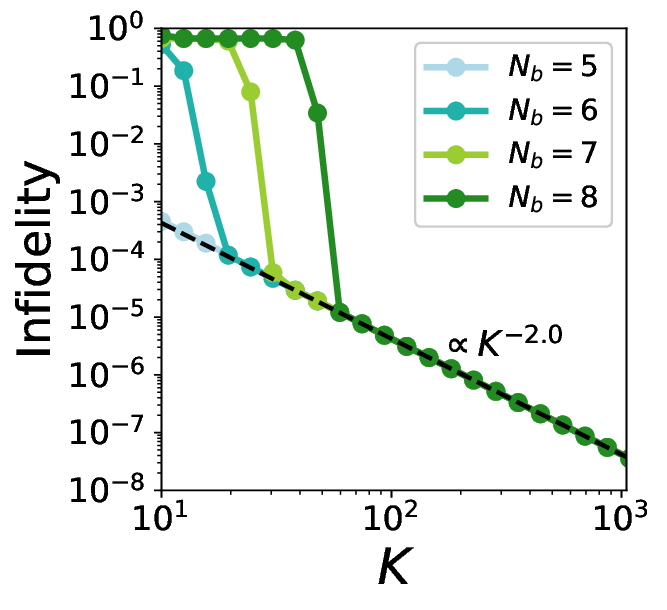}
    }
    \caption{Infidelity between the state $\ket{\Psi_K}$ -- which can be created using the method described in this Appendix -- and the encoded vacuum state $\ket{0_E}$ in Eq.~\eqref{eq:encoded-vacuum-bin}. The infidelity is plotted for $\delta = 1/\sqrt{2K}$ throughout. }
    \label{fig:infidelity_vs_Nb_and_K}
\end{figure*}

The vacuum state of a bosonic mode is given as 
\begin{align}
    \ket{m=0}=\frac{1}{\pi^{1/4}} \int_{-\infty}^{\infty}dx \,e^{-x^2/2}  \ket{x},
\end{align}
with variance $\langle m=0 | \,x^2 |m=0\rangle=1/2$.
An approximate vacuum state in the binary-x encoding (called vacuum Hermite state \cite{Somma1Dsystems, QuantumHermiteTransform} and note \footnote{The norm of the state $\ket{m=0}_E$ is exponentially (in $N_b$) close to 1.}) can be obtained by simply using the expression for the encoded quadrature operator $x_E=x_E(j)$ in Eq.~\eqref{eq:xbin}  
\begin{align}
    \ket{m=0_E} =\left(\frac{2\pi}{L}\right)^{1/4}\frac{1}{\pi^{1/4}}\sum_{j=0}^{2^{N_b}-1}\exp(-x_E^2(j)/2) \ket{j}.
    \label{eq:encoded-vacuum-bin}
\end{align}
Of course, many approximate states are possible, and we seek an approximation which is most-efficient to prepare as a quantum circuit.
The cost of implementing Gaussian states was considered in Ref.~\cite{kitaevwebb, Somma1Dsystems}, but more efficient methods using block encodings have been proposed in \cite{kuklinski2025} using $O(N_b^2)$ Toffoli gates, leading to 10-300 CNOT gates for $N_b\leq 9$.

In this work we opt for a post-selection method using ancilla qubits with controlled-displacements for which we can evaluate the CNOT count easily. Note that the Dicke model simulation also uses controlled-displacements and hence our method simply adds to those estimates. The method is similar to what is done experimentally to prepare a state with a wider Gaussian envelope from the vacuum state \cite{terhalEncodingQubitCavity2016,CampagneIbarcq2020}.

Starting from the state representing $x_E=0$, the method iteratively creates equal superpositions of states that are translated forwards and backwards in the x-direction, creating a binomial amplitude distribution which approximates a Gaussian distribution for a large number of iterations $K$. 
Consider first an idealized protocol on a bosonic mode. We start with a state $\ket{x=0}$ (a perfect eigenstate of the $x$ quadrature operator). Using an ancilla qubit in $\ket{+}$, we apply a qubit-controlled displacement $D(\delta)$ with $\delta \in \mathbb{R}$, depending on the ancilla qubit state. We do this for $K$ ancilla qubits and postselect on all ancilla qubits being measured in $\ket{+}$. For such evolution $U$, let
\begin{align}
B_{K,\delta}=\bra{+^{\otimes K}} U  \ket{+^{\otimes K}}=\frac{1}{2^K}(D(\delta)+D(-\delta))^K=
\frac{1}{2^K}\sum_{k=0}^K {K \choose k} D(\delta(K-2k)).
\end{align}

Using that $D(\delta)\ket{x=0}= e^{-i\sqrt{2} \delta p}\ket{x=0}=\ket{x=\sqrt{2}\delta}$, we have
\begin{align}
    B_{K,\delta}\ket{x=0}=\frac{1}{2^K}\sum_{k=0}^K {K \choose k}\ket{x=(K-2k)\sqrt{2}\delta}, \quad ||B_{K,\delta} \ket{x=0}||\geq \sqrt{\frac{{K \choose \lfloor K/2\rfloor}^2}{2^{2K}}} \geq \frac{1}{\sqrt{2(K+1)}},
    \label{eq:ideal-post}
\end{align}
where we have used $\binom{K}{\lfloor K/2\rfloor}\geq 2^{K}/\sqrt{2(K+1)}$.
The normalized state $\ket{\Psi_K}=\frac{B_{K,\delta}\ket{x=0}}{||B_{K,\delta}\ket{x=0}||}$ has variance 
$\sigma^2=\bra{\Psi_K} x^2 \ket{\Psi_K} \approx K \delta^2$. Since for the ideal vacuum state the quadrature variance is $1/2$, we set $\delta=\frac{1}{\sqrt{2K}}$. We take $K$ sufficiently large so that the binomial distribution approximates a Gaussian. Note that $N_b$ should not be chosen too large for a given $K$, since the distribution will then correspond to a sequence of resolved peaks as opposed to an approximate Gaussian, resulting in a small overlap between $\ket{\Psi_K}$ and $\ket{m=0}_{E}$. This is clearly visible in Figure~\ref{fig:AppBb} below. 

Now, let us discuss the encoded version of this protocol. Using Eq.~(\ref{eq:binaryxdisplacement}) we can decompose each qubit-controlled displacement in the $p$-direction as $V^{\dagger}\exp(-i \sqrt{2}\delta x_E)V$ where $V=Z_0 QFT Z_0$. We can thus do the operation $V$ once at the start of the protocol on the input state representing 
$x_E=0$, i.e. $\ket{j=2^{N_b-1}}=\ket{100\ldots 0}$, and apply $V^{\dagger}$ again at the very end, after all $K$ ancilla qubits have interacted. Thus the number of QFTs is two, independent of $K$. Besides the QFT and $Z_0$ gates, each ancilla uses a qubit-controlled $\exp(-i \sqrt{2}\delta x_E)$ operation. The implementation of this gate using the GHZ-trick has been discussed in Section~\ref{sec:mindepth}. Each ancilla qubit can interact with the bosonic register qubits sequentially. 


Assuming postselection, it follows that this method has a CNOT count of $2\#(QFT) + 2N_b K$ and depth $2D(QFT) + 3N_b K$ (or $2D(QFT) + K\big(2\lceil \log_2(N_b) \rceil + 3\big)$ when using $N_b$ additional ancilla qubits and employing the GHZ-trick (which we have available by assumption when doing simulations of the Dicke model), with $\#(QFT)$ and $D(QFT)$ the CNOT count and depth for a single QFT implementation and $K$ the number of iterations. The probability for success in Eq.~\eqref{eq:ideal-post} goes as $1/\sqrt{K}$. 

Note that in Fig.~\ref{fig:encoding_errors} and Fig.~\ref{fig:CNOTcount_displacements} in the main text, we use Eq.~\eqref{eq:encoded-vacuum-bin} to represent the vacuum state, hence only representing the encoding error. In all other Figures in the main text, the error induced by the vacuum preparation circuit implementation is simulated and hence included with $K$ specified.

For completeness, we plot the infidelity between the encoded vacuum state $\ket{0_E}$ in Eq.~\eqref{eq:encoded-vacuum-bin} and $\ket{\Psi_K}$ (with $\delta = \frac{1}{\sqrt{2K}}$) in Fig.~\ref{fig:infidelity_vs_Nb_and_K}. Clearly, the infidelity does not depend strongly on $N_b$ as long as $K$ is sufficiently large and decays as $\Theta(1/K^2)$ for such large $K$. Note that the minimum $K$ required to get to the $\Theta(1/K^2)$ scaling might scale exponentially in $N_b$. 


\section{Costings for Dicke Hamiltonian evolution: circuit decompositions and Trotter error bounds}
\label{app:circuitdecompDicke}

\begin{table*}[htb]
\centering
\renewcommand{\arraystretch}{1.3}
\resizebox{\textwidth}{!}{%
\begin{tabular}{||l|c|c|c|c||}
\hline 
Encoding & Circuit type & 
\makecell{Trotter Layer Depth} &
\makecell{Trotter layer \\ CNOT count} &
\makecell{Trotter error $\epsilon_T$} \\
\hline\hline
Unary Fock   & \makecell{1-qubit gates + \\ controlled displacements} & $1+N_a(2\lceil \log_2(N_b) \rceil + 28)$ & $10N_a(N_b-2)$ &  $\frac{3 gt^2 \sqrt{N_a}}{2 n_T^D} \max(2g \sqrt{N_a}N_b, \omega_0 \sqrt{N_b},2\omega)$  \\
Dicke      & \makecell{1-qubit gates + \\ controlled displacements} & $1+N_a(2\lceil \log_2(N_b) \rceil + 5)$ & $2N_a(2N_b-1)$ & $\frac{g t^2 \sqrt{N_a N_b}}{n_T^D} \max(\omega, \omega_0)$  \\
Binary quadrature & \makecell{1- and 2-qubit gates + \\ controlled displacements} & \makecell{$14N_b-9+$ \\ $N_a(2\lceil \log_2(N_b) \rceil + 5) \:^{*}$} & \makecell{$2N_b^2+2N_a(2N_b-1) \:^{*}$} & $\frac{3 t^2 \sqrt{\pi 2^{N_b}}}{2n_T^D}\max(\pi^{3/2} \omega^2 2^{3N_b/2},g\omega \pi\sqrt{N_a}2^{N_b},g\omega_0 \sqrt{N_a})$  \\
\hline
\end{tabular}%
}
\caption{Comparison between encodings in terms of CNOT count and depth for evolution under the Dicke Hamiltonian using $N_b$ (bosonic) qubits and $N_a$ atom-qubits. The costs are shown for a single Trotter layer. The Trotter error is given by $||U_E(t)-\tilde{U}_E(t)||_E\leq \epsilon_T$ where $\tilde{U}_E(t)$ is the encoded Trotterized evolution for time $t$, and $U_E(t)$ is the encoded evolution for time $t$ (hence it does not include the encoding error), using $n_T^D$ Trotter layers. We evaluate the norm within the codespace itself, hence the notation $||\cdot||_E$. The circuit depth is obtained by decomposing the Trotterized circuits into CNOT and single-qubit gates, using parallelizations as described in the main text, and using at most $N_b$ ancillary qubits at the time. We do not apply optimizations such as merging single-qubit gates after this decomposition.}
\label{tab:encoding-comparison}
\end{table*}

We bound the Trotter error in the encoded Dicke model time evolution using Eq.~\eqref{eq:Hdicke_trotter} and the encoded Hamiltonians $H_{a,E},H_{b,E},H_{ab,E}$, to fill in the last column of Table~\ref{tab:encoding-comparison}. These Trotter errors are not used in any numerical results in this paper ---the actual Trotterized circuits are classically simulated--- but they serve as a comparative scaling assessment. However, due to some bounds being looser than others, no clear comparative conclusions can be drawn from them.
  
Similar as for displacements (Appendix~\ref{app:Trotter_derivation}), we bound the Trotter error within the encoded space and explicitly denote this as $||.||_E$. Given the circuit decomposition of the Trotterized evolution, Section~\ref{sec:res_est} has described how we Trotterize and parallelize the circuit implementation using ancillary qubits. Paulis with superscript $a$ (resp. $b$) act on atom-qubits (resp. bosonic qubits representing the bosonic mode). The Trotterization and use of QFT circuit (binary quadrature encoding) generally provides a circuit with two-qubit gates of the general form $\exp(-i \theta P_{k} P_j)$ (Dicke, binary and quadrature encoding), or also some three-qubit gates of the general form  $\exp(-i \theta P_k^b P_{k+1}^b X_j^a)$ (unary encoding). Such gates are then enacted using CNOT gates and single-qubit gates in a standard manner \cite{book:NC}, e.g.
\begin{equation}
    \exp(-i \theta P_{k}P_j)= V_1CNOT_{j\to k}\exp(-i\theta Z_{k})CNOT_{j\to k}V_1^{\dagger},
\end{equation}
with $V_1$ some single-qubit gates, thus using 2 CNOT gates and a circuit of total depth 5. Similarly, a 3-qubit Pauli rotation uses 4 CNOTs and a circuit of total depth 7. These costings result in the columns on CNOT count and circuit depth in Table~\ref{tab:encoding-comparison}.


\subsection{Dicke encoding}
\label{app:dicke}

We use the bound in Eq.~\eqref{eq:trotter_error} and triangle inequalities for operator norms to bound the Trotter error for the Dicke encoding. Since $H_{b,E}$ and $H_{a,E}$ commute, we only have to evaluate the norms of $[H_{b,E},H_{ab,E}]$ and $[H_{a,E},H_{ab,E}]$. 
Using Eq.~\eqref{eq:anni-Dicke}, one can readily show
\begin{align}
[H_{b,E},H_{ab,E}]&=\frac{i g\omega_0}{\sqrt{N_b N_a}}\sum_{j=0}^{N_b-1} Y_j^b \otimes \left(\sum_{k=1}^{N_a} X^a_k\right).
\end{align}
The norm is achieved by putting each atom-qubit in $\ket{+_Y}$ and each bosonic qubit in $\ket{+}$, leading to 
\begin{align}
\lVert [H_{b,E},H_{ab,E}] \rVert_E =g\omega_0 \sqrt{N_bN_a} \label{eq:estimate_gk_app}.
\end{align}
Analogously, we obtain $ \lVert [H_{a,E},H_{ab,E}] \rVert_E = g\omega \sqrt{N_bN_a}$. Hence, we get
\begin{equation}
    \epsilon_T =\frac{t^2}{2n_T^D}\left[g\omega \sqrt{N_bN_a}+g\omega_0 \sqrt{N_bN_a}\right].
\end{equation}


\subsection{Unary Fock encoding}

For the unary encoding, we Trotterize using Eq.~\eqref{eq:Hdicke_trotter} and then additionally Trotterize $e^{-iH_{ab,E}t/n_T^D}$ using Eq.~\eqref{eq:displacement_unary_trotter} for controlled-displacements. That is, we can write
\begin{align}
    U_E(t)=\exp(-i H_E t) \rightarrow  \tilde{U}_E(t)=\left(e^{-i t H_{b,E}/n_T^D}e^{-i t H_a/n_T^D} e^{-i t H_{ab,E}^e/n_T^D}e^{-i t H_{ab,E}^o/n_T^D} \right)^{n_T^D}
    \label{eq:evenodd}
\end{align}
where $H_{ab,E}=H_{ab,E}^o+H_{ab,E}^e$ with
\begin{align}
 H_{ab,E}^e= \frac{g}{\sqrt{N_a}} \sum_{j=1}^{N_a} X^a_j \otimes H_E^e(z=i), \notag \\
 H_{ab,E}^o= \frac{g}{\sqrt{N_a}}\sum_{j=1}^{N_a} X^a_j \otimes H_E^o(z=i),
\end{align}
with $H_E^{e/o}(z)$ defined in Eqs.~\eqref{eq:displacement_unary_He},\eqref{eq:displacement_unary_Ho}. Due to Eq.~\eqref{eq:trotter_error}, we have
\begin{align}
     \lVert U_E(t)-\tilde{U}_E(t) \rVert_E \leq \frac{t^2}{2n_T^D} \left(||[H_{a,E},H_{ab,E}]||_E+||[H_{b,E},H_{ab,E}]||_E+||[H_{ab,E}^e,H_{ab,E}^o]||_E\right).
\end{align}
Using Eq.~\eqref{eq:boundA} we have
\begin{align}
  ||[H_{ab,E}^o,H_{ab,E}^e]||_E\leq 2 g^2 N_a N_b.
  \label{eq:evenodd}
\end{align}
and
\begin{align}
    ||[H_{a,E},H_{ab,E}]||_E\leq \omega_0 g \sqrt{N_a} ||H_{E}(z=i)||_E\leq \omega_0 g \sqrt{N_a N_b},
\end{align}
and 
\begin{align}
    ||[H_{b,E},H_{ab,E}]||_E\leq \omega g \sqrt{N_a}/2 ||[\sum_{j=0}^{N_b-1} j Z_j,\sum_{k=0}^{N_b-2} \sqrt{k+1} \sigma_{k+1}^-\sigma_k^++h.c.]]||_E\leq 2\omega g \sqrt{N_a}.
    \label{eq:upp}
\end{align}
We note that the upperbound in Eq.~\eqref{eq:upp} goes to $O(N_b)$ if we evaluate the norm of the full space of $N_b$ qubits not only on the encoded subspace. Thus we get
\begin{align}
    \epsilon_T = \frac{t^2}{2 n_T}\left[2\omega g \sqrt{N_a}+\omega_0 g \sqrt{N_a N_b}+2 g^2 N_a N_b\right].
\end{align}
Clearly, in the superradiant regime when $g> g_c=\sqrt{\omega \omega_0}/2$ and for large $N_a$ versus $N_b$, the commutator norm, obtained by Trotterizing the coupling term, in Eq.~\eqref{eq:evenodd} dominates. The dependence on $N_b$ worsens to $O(N_b^2)$ when we evaluate the norms on the full space.


\subsection{Binary quadrature encoding}

For the binary quadrature encoding, we again use Eq.~\eqref{eq:Hdicke_trotter}. For the free bosonic evolution with $H_{b,E}$ we use a simple Trotterized form:
\begin{align}
\exp(-i t \omega a^{\dagger}_E a_E/n_T^D) \rightarrow \exp(-i \omega t p_E^2/(2 n_T^D))  \exp(-i \omega t x_E^2/(2 n_T^D)). 
\label{eq:trotter-free}
\end{align}
Note that we could have used the fast-forwarding decomposition in Eq.~\eqref{eq:ff} but due to small $t/n_T^D$ the difference in numerical practice is expected to be small: basically, by Taylor expanding $a(t)$ and $b(t)$ to first order in $t$ in Eq.~\eqref{eq:ff} one gets a second-order (instead of a first-order) Trotterization of the evolution.

Thus we use that $\frac{\omega}{2} [x_E^2,H_{ab,E}]=0$ and evaluate the commutators $\frac{\omega}{2} [p_E^2,H_{ab,E}]$, $\frac{\omega^2}{4}[p_E^2,x_E^2]$ and $[H_{a,E},H_{ab,E}]$ with 
\begin{align}
     H_{ab,E}=\frac{g}{\sqrt{N_a}}\sum_j X_j \otimes \left(\sqrt{2}x_E \right).\label{eq:H_ab_binary_x_app}
\end{align}
Since the full space of the $N_b$ qubits is the encoded space for binary-quadrature encoding, we drop the $_E$ labels.
We have
\begin{align}
||[[H_{a,E},H_{ab,E}]]||=g\omega_0  \sqrt{2 N_a}||x_E||\leq g\omega_0  \sqrt{\pi} \sqrt{N_a 2^{N_b}}.
\end{align}
and, using $||x_E||=||p_E||=\sqrt{\pi}\sqrt{2^{N_b-1}}$,
\begin{align}
\frac{\omega}{2}||[p_E^2,H_{ab,E}]||=g\omega  \sqrt{N_a/2} ||[p^2_E,x_E]||\leq \omega g \sqrt{2 N_a}   ||p_E|| \times ||[[p_E,x_E]]|| =g\omega  \sqrt{\pi}\sqrt{N_a2^{N_b}}  \times ||[[p_E,x_E]]||.\notag \\
 \frac{\omega^2}{4}||[p_E^2,x_E^2]||\leq \frac{\omega^2}{2}||x_E|| \times ||p_E|| \times ||[p_E,x_E]||\leq \omega^2 \pi 2^{N_b} ||[p_E,x_E]||.
\end{align}
The commutator norm $||[[p_E,x_E]]||$ can be bounded by bounding the norm of the centrally-shifted operators $p_E$ and $x_E$ themselves, i.e. 
\begin{align}
   ||[p_E,x_E]||\leq \frac{2\pi}{2^{N_b}}(2^{N_b}-1)^2/2\leq \pi 2^{N_b}.
\end{align}
Summing everything up gives a bound
\begin{align}
    \epsilon_T= \frac{t^2}{2n_T^D}\left[\pi^2 \omega^2 2^{2N_b}+g \omega \pi\sqrt{\pi}\sqrt{N_a}2^{3N_b/2}+g\omega_0  \sqrt{\pi}\sqrt{N_a}2^{N_b/2}\right].    
\end{align}
For small $N_a$ versus $2^{N_b}$ this (worst-case) bound is dominated by the Trotterization in Eq.~\eqref{eq:trotter-free}, which is likely overly-pessimistic as the upperbound on $||[[p_E,x_E]]||$ is unlikely to capture average behavior.

\section{Supplementary information about the H2-2 Quantinuum device experiments} \label{app:supp_info}

We implemented all experiments using pytket version 2.16.0~\cite{Pytket,Sivarajah_2021} and qnexus 0.39.0~\cite{quantinuum_nexus}. The circuits ran on Quantinuum's 56 trapped-ion qubit device H2-2 (version 0.1.0). The physical location of the qubits are adjusted so that two-qubit gates are performed in one of four two-qubit gate-interaction zones, allowing for all-to-all connectivity~\cite{quantinuum_h22} in principle. We list the provided error data of the device at the time of the experiment (May 2026) in Table~\ref{tab:h2-2-errors}.

\begin{table}[ht]
\centering
\begin{tabular}{lcc}
\hline
\textbf{Error source} & \textbf{Error probability} & \textbf{Uncertainty} \\
\hline
1-qubit gate error       & $2.8\times10^{-5}$ & $3.6\times10^{-6}$ \\
2-qubit gate error       & $8.3\times10^{-4}$ & $4.8\times10^{-5}$ \\
Crosstalk error          & $2.2\times10^{-5}$ & $5.3\times10^{-7}$ \\
Memory error             & $1.2\times10^{-4}$ & $2.0\times10^{-5}$ \\
Measurement error ($|0\rangle$) & $6.7\times10^{-4}$ & $8.7\times10^{-5}$ \\
Measurement error ($|1\rangle$) & $1.2\times10^{-3}$ & $1.1\times10^{-4}$ \\
\hline
\end{tabular}
\caption{Error parameters reported for the Quantinuum H2-2 model. 
The values correspond to the noise snapshot dated 28.08.2025. 
Uncertainties are reported by Quantinuum.}
\label{tab:h2-2-errors}
\end{table}

\subsection{Compiled depth comparison}
\label{app:opti}

\begin{figure}[htbp]
    \centering
    \subfloat[Unary encoding]{
        \includegraphics[width=0.4\linewidth]{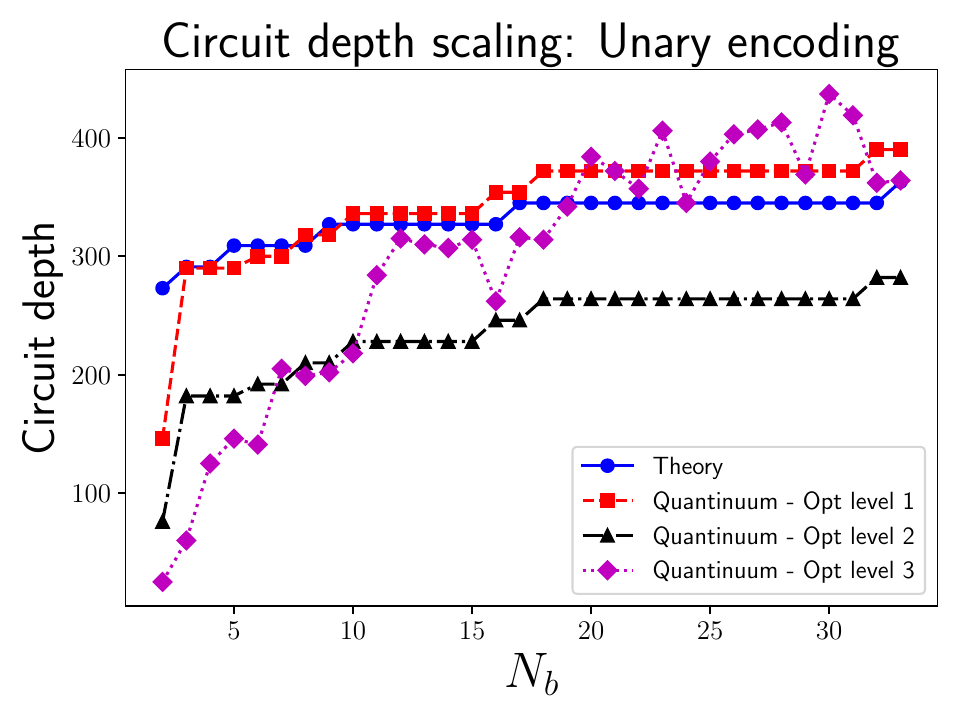}}
        \label{fig:unary_and_dicke_time_depth}
    \vspace{0.2cm}
    \subfloat[Dicke encoding.]{\includegraphics[width=0.4\linewidth]{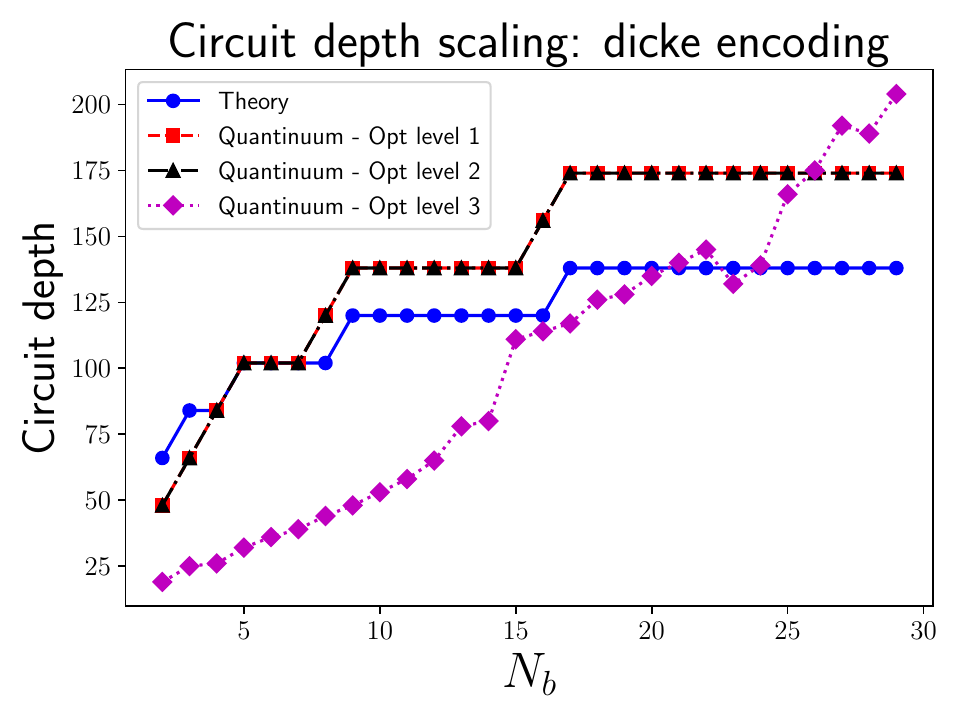}}
        \label{fig:trotter_comparison_time_depth}
    \caption{Depth scaling of the compiled circuits implementing the time evolution of the Dicke Hamiltonian for different pytket optimisation levels. The number of atom-qubits is $N_a = 3$ and we chose more than one Trotter layer to give the compiler the possibility to simplify the gate structure generated by two successive Trotter layers. The parameters $\omega$, $\omega_0$, coupling strengths $g$, the time duration are non-zero. We do not further specify them, as their exact values have no impact on the depth. The number of Trotter layer $n_T^D$ was set to three for both encodings. The blue curves with circle markers was obtained using the expression for circuit depth in Table~\ref{tab:encoding-comparison}. (a) Data for unary Fock encoding. (b) Data for Dicke encoding.}
    \label{fig:depth_comparison}
\end{figure}

Pytket compiled the circuits for the H2-2 device. That is, it maps an input circuit to a circuit respecting the connectivity and native gate set of the respective hardware and ---depending on the optimization level--- optimizes the circuit's gate ordering and composition. It also adapts the circuit such that it is compatible with the required re-arrangement of qubits into the interaction zones. For the compilation one can choose optimization levels 0-3. Due to the large size of the compiled circuits and lack of human readability, an in-depth analysis of compiler behavior is infeasible. Instead, we have analysed the circuit depth scaling when varying $N_b$ for all optimization levels and compared them to the theoretically expected scaling as given in Table \ref{tab:encoding-comparison}, see Fig.~\ref{fig:depth_comparison}. Here, the optimisation is such that it can never use more than $N_a+2N_b$ qubits. 

There are several factors which can influence the different depth estimates seen in Fig.~\ref{fig:depth_comparison}.
First, for small $N_b$ and $N_a$, our depth counts in Table \ref{tab:encoding-comparison} are worse than need-be as the GHZ trick can be replaced by an edge coloring scheme from Section \ref{sec:edge}. Second, the device has limited two-qubit gate parallelism, i.e. 4 pairs of qubits can do a two-qubit gate at the time, while we assume arbitrary parallelism. Third, we do not try to merge single-qubit gates together or lump single-qubit gates with two-qubit gates, and we compile to CNOT gates, a common standard. On the device, the compilation is to a continuously-parametrized ZZ-gate. 

We observe that optimisation level 3 finds smaller circuit depths for small $N_b$, while seemingly performing worse for larger $N_b$. It is beyond the scope of this paper to analyse the inner mechanisms of the pytket compiler and thus we treat it as a black box here. As we choose a rather small $N_b$ for the time evolution for both encodings, we pick optimisation level 3 to minimise the running costs. However, for experiments on a larger scale, we recommend choosing optimisation level 2 due to the seemingly worse behavior of the compiler for optimisation level 3. 

\subsection{Error analysis}\label{sec:supp_info_error}
Here, we discuss the generation of confidence intervals for the plots in Fig.~\ref{fig:unary_and_dicke} and Fig.~\ref{fig:unary_and_dicke_time}. We first discuss how the confidence interval of the expectation $\langle p_E \rangle$ can be estimated from the confidence intervals of $p_{e,E}$ and $p_{o,E}$.
To estimate the confidence interval of $p_E=p_{e,E}+p_{o,E}$ we use the union bound. Let $c_e$ (resp. $c_o$) be the probability that $p_{e,E}$ (resp. $p_{o,E}$) is not in the interval $[a_e,b_e]$ (reps. $[a_o,b_o]$). Then, per the union bound, the probability that $p_E$ is not within $[a_e+a_o,b_e+b_o]$ is upper bounded by $c_e+c_o$, which means $\mathbb{P}(a_e+a_o \leq p_E \leq b_e+b_o)  \geq 1-c_e-c_o$. We thus first compute the $97.5\%$ confidence intervals of $p_{E,e}$ and $p_{E,o}$ to obtain the $95\%$ confidence interval of $p_E$. 

Due to limited access to the device, our sample sizes for the experiments described in Sections~\ref{sec:experiment_displacement} and ~\ref{sec:experiment_timeevo} had to be capped at $M_{\rm samp}=500$ or 1000 shots. The most rigorous approach entirely agnostic to any experimental conditions would be to repeat the entire experiment (i.e. all shots) on different days and then compute the variances of the estimated means over this full sample. Another way would be to derive a worst-case bound on the maximum possible variance based on the known expressions for the encoded number and momentum operators. Both methods required larger sample sizes than what was possible within our budget. Thus, we estimate confidence intervals for the estimated means in Fig.~\ref{fig:unary_and_dicke} and~\ref{fig:unary_and_dicke_time} by boot-strapping~\cite{bootstrap}. That is, we compute a distribution of values per shot for $N=\langle a^{\dagger} a \rangle(t)$ for a given time step $t$. The probability distribution $\mathbb{P}(N=x)$ is then obtained as a histogram with measured values for $N$ given $M_{\rm samp}$ shots. We then resample 1000 times from the estimated (boot-strapped) probability distribution (i.e. the histogram) and compute the confidence intervals from the obtained statistics. For displacement simulations, we proceed analogously for $p_{E}$, also using 1000 resamples. 
\end{document}